\documentclass[aps,amsmath,amssymb,twocolumn,longbibliography,floatfix]{revtex4-1}

\usepackage{mathrsfs} 
\usepackage{graphicx,color} 
\usepackage{mathtools}
\usepackage{bigints}
\usepackage{bm}
\usepackage{threeparttable}
\usepackage[normalem]{ulem}
\usepackage{physics}
\usepackage{color}
\usepackage[colorlinks=true]{hyperref}
\hypersetup{
  colorlinks=true,
  linkcolor=blue,
  filecolor=magenta,
  citecolor=blue,
  urlcolor=blue,
}

\newcommand{\blue}[1]{\textcolor{black}{#1}}

\begin{document}

\title{Elastic heterogeneity governs asymmetric adsorption-desorption in a soft porous crystal}

\author{Kota Mitsumoto}
\email{kmitsu@iis.u-tokyo.ac.jp}
\author{Kyohei Takae}%
\email{takae@iis.u-tokyo.ac.jp}
\affiliation{Department of Fundamental Engineering, Institute of Industrial Science, University of Tokyo, 4-6-1 Komaba, Meguro-ku, Tokyo 153-8505, Japan}

\date{\today}

\begin{abstract}
Metal--organic frameworks (MOFs), which possess a high degree of crystallinity and a large surface area with tunable inorganic nodes and organic linkers, exhibit high stimuli-responsiveness and molecular adsorption selectivity that enable various applications. The adsorption in MOFs changes the crystalline structure and elastic moduli. Thus, the coexistence of adsorbed/desorbed sites makes the host matrices elastically heterogeneous. However, the role of elastic heterogeneity in the adsorption--desorption transition has been overlooked. Here we show the asymmetric role of elastic heterogeneity in the adsorption--desorption transition. We construct a minimal model incorporating adsorption-induced lattice expansion\blue{/contraction} and an increase\blue{/decrease} in the elastic moduli.
We discover that the transition is hindered \blue{by the entropic and energetic effects which become asymmetric in adsorption process and desorption process},
leading to the strong hysteretic nature of the transition. Furthermore, the adsorbed/desorbed sites exhibit spatially heterogeneous domain formation, implying that the domain morphology and interfacial area between adsorbed/desorbed sites can be controlled by elastic heterogeneity. Our results provide a theoretical guideline for designing soft porous crystals with tunable adsorption hysteresis and the dispersion and domain morphology of adsorbates using elastic heterogeneity.
\end{abstract}

\maketitle

Nature utilizes the mechanical flexibility of porous materials to control their functionality.
For example, liquid-liquid phase separation is suppressed by elastic stress exerted by polymer networks, where the elastic modulus of the network is an important factor determining nucleation temperature and droplet size~\cite{rosowski2020elastic}. This suppression is crucial in biological cells, where the viscoelastic chromatin network slows down the formation and diffusion of biomolecular condensates~\cite{lee2021chromatin}. Molecular adsorption in soft porous materials is also crucial in industrial applications.
Metal--organic frameworks (MOFs) possess controllable mechanical flexibility and pore size due to a high degree of crystallinity and a large surface area with tunable inorganic nodes and organic linkers~\cite{horike2009soft,furukawa2013chemistry}. Their mechanical flexibility is utilized to control stimuli-responsiveness and molecular adsorption selectivity that enables diverse applications, including gas separation, storage, and release~\cite{li2012metal}, sensors~\cite{kreno2012metal}, biomedicines and enzyme protection~\cite{horcajada2012metal,liang2021metal}, catalysts~\cite{bavykina2020metal}, supercapacitors~\cite{sheberla2017conductive}, and actuators~\cite{terzopoulou2020metal}. From a thermodynamic perspective, the mechanical deformation of MOFs is governed by their elastic moduli and their dependencies on the pressure, temperature, and adsorption of gas/solvent molecules~\cite{Landau7}. Therefore, elucidating the physical mechanisms connecting the macroscopic elasticity to the microscopic interaction of MOFs is of growing interest to develop guidelines for designing MOFs with desired chemo-mechanical functions.

Because adsorbed molecules strongly interact with inorganic nodes and organic linkers, the crystalline structure and elastic moduli change upon molecular adsorption~\cite{horike2009soft,coudert2015responsive}. Although there have been extensive studies on structural transformation, its connection with the controllability of the adsorption transition remains elusive. There have also been studies on changes in elastic moduli upon homogeneous molecular adsorption~\cite{ortiz2013investigating,henke2014guest,mouhat2015softening,canepa2015structural}, focusing on the mechanical stability of MOFs. However, their effect on the transition has not been investigated. Furthermore, adsorbed molecules are distributed heterogeneously in MOFs~\cite{cho2015extra,rogge2019unraveling}, implying that the local lattice constant and elastic moduli of the substances become heterogeneous even without the disorder, depending on the spatial distribution of the adsorbates. The difference in the local lattice constant between the adsorbed and desorbed sites induces a lattice mismatch. Other elastically heterogeneous systems, such as phase separating alloys, martensites, ferromagnets with elastic coupling, and spin-crossover solids, also exhibit the lattice mismatch. In such systems, the lattice mismatch is relaxed by the emergence of heterogeneous lattice distortion reflecting the local elastic moduli, affecting the phase behavior, transformation kinetics, and domain morphology~\cite{Khachaturyan,Onukibook,mitchell2006phase,bousseksou2011molecular}. Similarly, it is crucial to reveal the role of elastic heterogeneity in MOFs. Because elastic interactions have a long-range nature~\cite{Landau7}, a large system size is required in numerical simulations to examine the role of elastic heterogeneity; thus, quantum-chemical calculations~\cite{odoh2015quantum} and {\it ab initio} molecular dynamics simulations~\cite{coudert2016computational} require vast computational resources and costs. Furthermore, data-driven approaches~\cite{jablonka2020big} are currently inefficient because data on adsorption-induced changes in the elastic moduli are scarce. Therefore, a coarse-grained molecular model needs to be studied~\cite{watanabe2009free,mouhat2015softening,enachescu2018elastic} to elucidate the long-range nature of elastic heterogeneity with manageable computational costs.

In this study, we elucidate the role of elastic heterogeneity in the adsorption/desorption of guest molecules in a soft porous framework. By constructing a minimal model incorporating the adsorption-induced lattice expansion\blue{/contraction} and hardening\blue{/softening}, we reveal that elastic heterogeneity exhibit different roles in gas adsorption and desorption. 
\blue{We mainly study the adsorption-induced lattice expansion and hardening case as a prototypical example.}
In the gas adsorption, the shape of the adsorbed domains is limited to isotropic shapes, reducing the entropy of the growing domains. In contrast, softer desorbed domains are flattened in the gas desorption to reduce elastic energy, increasing the interfacial energy. Thus, the growth of the adsorbed/desorbed domains is entropically/energetically hindered, leading to a robust hysteresis. 
\blue{This result is qualitatively unchanged in other cases: hysteretic adsorption-desorption transition occurs because the domains with larger/smaller elastic modulus have isotropic/anisotropic shapes, respectively.}
Our findings shed light on the utilization of elastic heterogeneity for capturing and distributing guest molecules.

\begin{figure}[t]
\centering
\includegraphics[width=85mm]{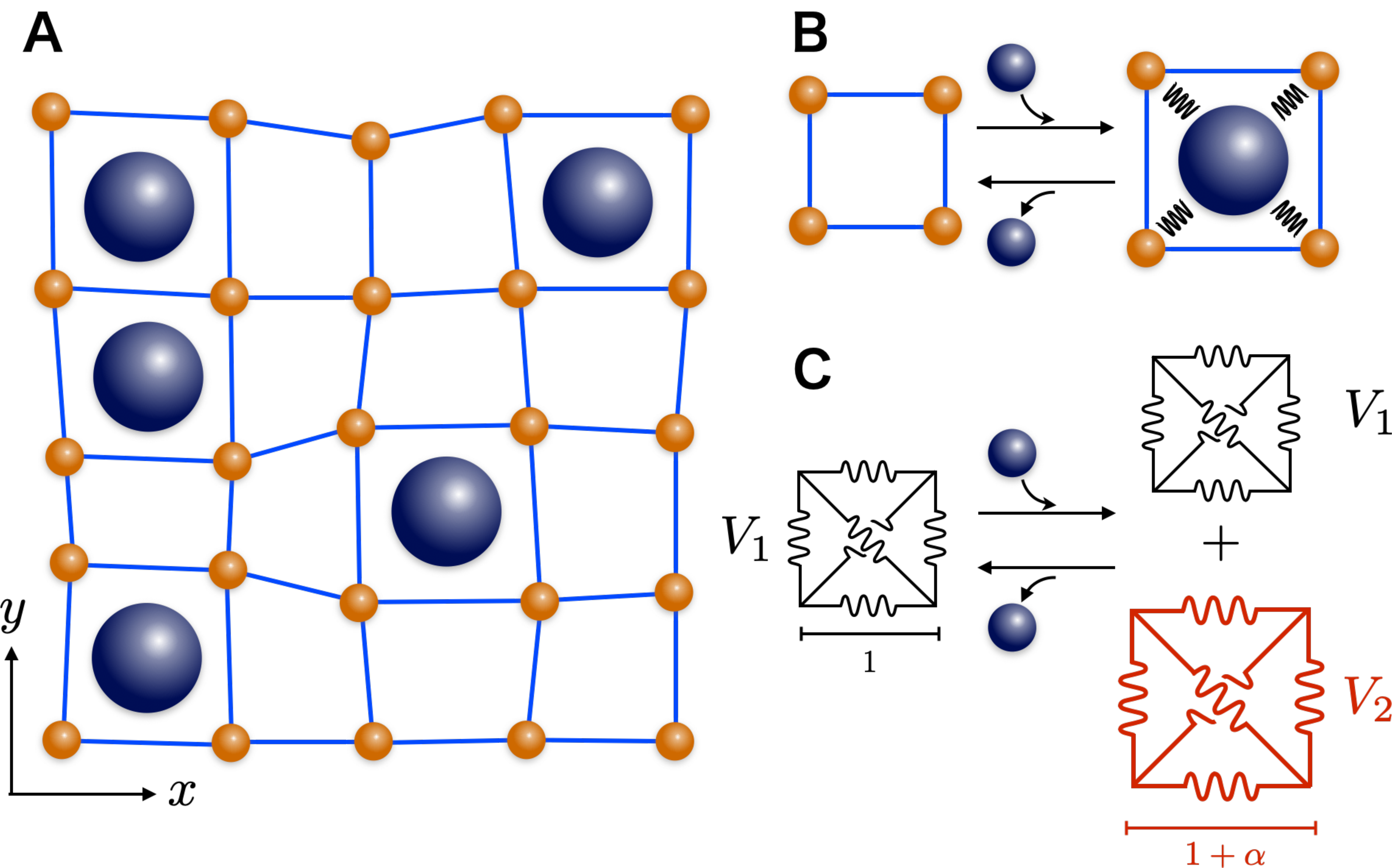}
\caption{
Schematic description of the proposed model.
({\it A}) Guest particles (dark blue spheres) are adsorbed/desorbed into the host matrix formed by the host particles (orange), inducing isotropic swelling/contraction of the host matrix locally.
({\it B}) An adsorbed particle strongly interacts with the \blue{host} particles.
({\it C}) Mathematical representation of {\it A} and {\it B}. Interaction potential $V_2$ arises from the filling of the guest particle in addition to the original host's potential $V_1$.
}
\label{fig:schematic}
\end{figure}

\section*{Results}
\subsection*{Minimal model incorporating elastic heterogeneity}
We construct a two-dimensional square-lattice model, as shown in Fig.~\ref{fig:schematic}{\it A}. As discussed below, this simplified model can capture the essential role of elastic heterogeneity. A guest particle is adsorbed and desorbed in a flexible host matrix, inducing lattice displacement due to the interaction between the host and guest particles (Fig.~\ref{fig:schematic}{\it B}). Each plaquette (unit cell) can accommodate only one guest particle, and the plaquette expands isotropically by adsorbing the guest particles. This process is mathematically expressed in Fig.~\ref{fig:schematic}{\it C}. Each host particle interacts with the nearest-neighbor (NN) and next-nearest-neighbor (NNN) sites such that the square lattice configuration is stable.
In fig.~\ref{fig:schematic}{\it C}, the interaction with NN (NNN) sites is noted by four horizontal and vertical (two diagonal) springs. The NN and NNN potentials of distance $r$ are $\frac{k_0}{2}(\ell_0-r)^2$ and $\frac{k_0}{2}(\sqrt{2}\ell_0-r)^2$, respectively, with elastic constant $k_0$ and natural length of the square plaquette $\ell_0$.
\blue{For simplicity, the NN and NNN interactions have the same spring constant $k_0$ (see also Discussion).}
Hereafter, all quantities are expressed in units of $\ell_0$ (length unit) and $k_0\ell_0^2$ (energy unit). Thus, the potential energy of a plaquette $\square$ reads $V_1(\{\bm{r}_{i \in \square} \})=\frac{1}{4}\sum_{\rm NN}(1-r_{ij})^2 + \frac{1}{2}\sum_{\rm NNN}(\sqrt{2}-r_{ij})^2$, where $\{\bm{r}_{i \in \square} \}$ represents the position of the lattice sites at the vertices of the plaquette $\square$.
The interaction between the host and guest particles is incorporated as additional potential energy $V_2(\{\bm{r}_{i \in \square} \})=k[\frac{1}{4}\sum_{\rm NN}(1+\alpha-r_{ij})^2 + \frac{1}{2}\sum_{\rm NNN}(\sqrt{2}(1+\alpha)-r_{ij})^2]$, where $k$ is the relative energy scale of the host-guest interaction, and $\alpha$ represents the amplitude of the plaquette deformation. Thus, when guest particles are adsorbed homogeneously, the equilibrium lattice constant and rigidity become $1+k\alpha/(1+k)$ and $1+k$, respectively.
\blue{If $k\alpha>0$, the adsorption of the guest particles induces lattice expansion. This has been observed for various MOFs, including MIL-88~\cite{serre2007role}, narrow pore (np) to large pore (lp) transition in MIL-53 (Cr)~\cite{ferey2009large}, and gate-opening-type MOFs~\cite{kitaura2003porous}. In contrast, if $k\alpha<0$, adsorption induces the contraction of the matrix. This has also been observed experimentally in open pore to closed pore transition in DUT-49~\cite{krause2016pressure}, [Zn$_2$(terephthalate)$_2$(triethylenediamine)]$_n$ upon accommodation of {\it trans}-azobenzene~\cite{yanai2012guest}, and a uranium MOF~\cite{halter2020self}. Moreover, COF-300 expands (contracts) when tetrahydrofuran (water) molecules are adsorbed~\cite{chen2019guest}. Thus, the sign of $k\alpha$ depends on the substance and adsorbate. In the following, we mainly focus on the case of $k\alpha>0$ and $k>0$, such that the lattice expands and becomes more rigid by particle adsorption. Contrary to the swelling/contraction studies, those on adsorption-induced hardening/softening are scarce. Although several studies have reported adsorption hardening for zeolites~\cite{coasne2011enhanced}, MOF-74-Zn~\cite{canepa2015structural}, ZIF-4 (to be precise, softening upon evacuation of the framework)~\cite{bennett2011reversible}, and ZIF-8~\cite{ortiz2013investigating}, and softening for MIL-53 (Cr) by the np to lp transition~\cite{neimark2011structural}, and model microporous materials~\cite{mouhat2015softening}, the role of adsorption hardening/softening on the adsorption-desorption transition has been overlooked (see Discussion for estimated values of $k$ and $\alpha$ for some MOFs).}

Because we focus on isotropic swelling by particle adsorption, it is sufficient to adopt an osmotic ensemble~\cite{coudert2008thermodynamics} where the hydrostatic pressure $P$ is controlled instead of the anisotropic stress tensor. The other control parameters in this ensemble are the temperature $T$, the chemical potential of the guest particle adsorption $\mu_{\rm ads}$, and the number of particles in the host framework $N_{\rm host}$. The osmotic grand potential is defined as $\Omega=U-TS+PV-\mu_{\rm ads}N_{\rm ads}$, where $U$ is the energy, $S$ is the entropy, $V$ is the volume, and $N_{\rm ads}$ is the number of the adsorbed particles. Hence, its differential form reads
\begin{equation}
d\Omega = -SdT + VdP - N_{\rm ads}d\mu_{\rm ads} + \mu_{\rm host}dN_{\rm host},
\label{eq:thermo}
\end{equation}
where $\mu_{\rm host}$ is the chemical potential of the host particles. In this study, we fix $P=0$ and $N_{\rm host}=L\times L$, where $L$ is the linear system size of the square lattice. Hereafter, we denote $\mu_{\rm ads}$ and $N_{\rm host}$ as $\mu$ and $N$, respectively.
The Hamiltonian $\cal H$ comprises the lattice site positions $\bm{r}_i~(i=1,2,...,N)$ and guest variables on the plaquettes $\sigma_\square~(\square = 1,2,...,N)$ taking $1$ (presence) or $0$ (absence).
Thus,
\begin{equation}
{\cal H}=\sum_{\square = 1}^N \Big[V_1(\{\bm{r}_{i \in \square} \}) + \sigma_\square \qty[V_2(\{\bm{r}_{i \in \square} \})-\mu]\Big].
\label{eq:hamiltonian}
\end{equation}
We impose periodic boundary conditions in $x$ and $y$ directions with variable systems size to examine the bulk properties without the influence of the surface (see {\it Materials and Methods} for details).
We perform standard Monte Carlo (MC) simulations to examine transitions with hysteretic behavior, and multicanonical MC simulations using the Wang-Landau (WL) method to study the equilibrium phase transition (see {\it Materials and Methods} for details). 

\begin{figure}[t]
\centering
\includegraphics[width=85mm]{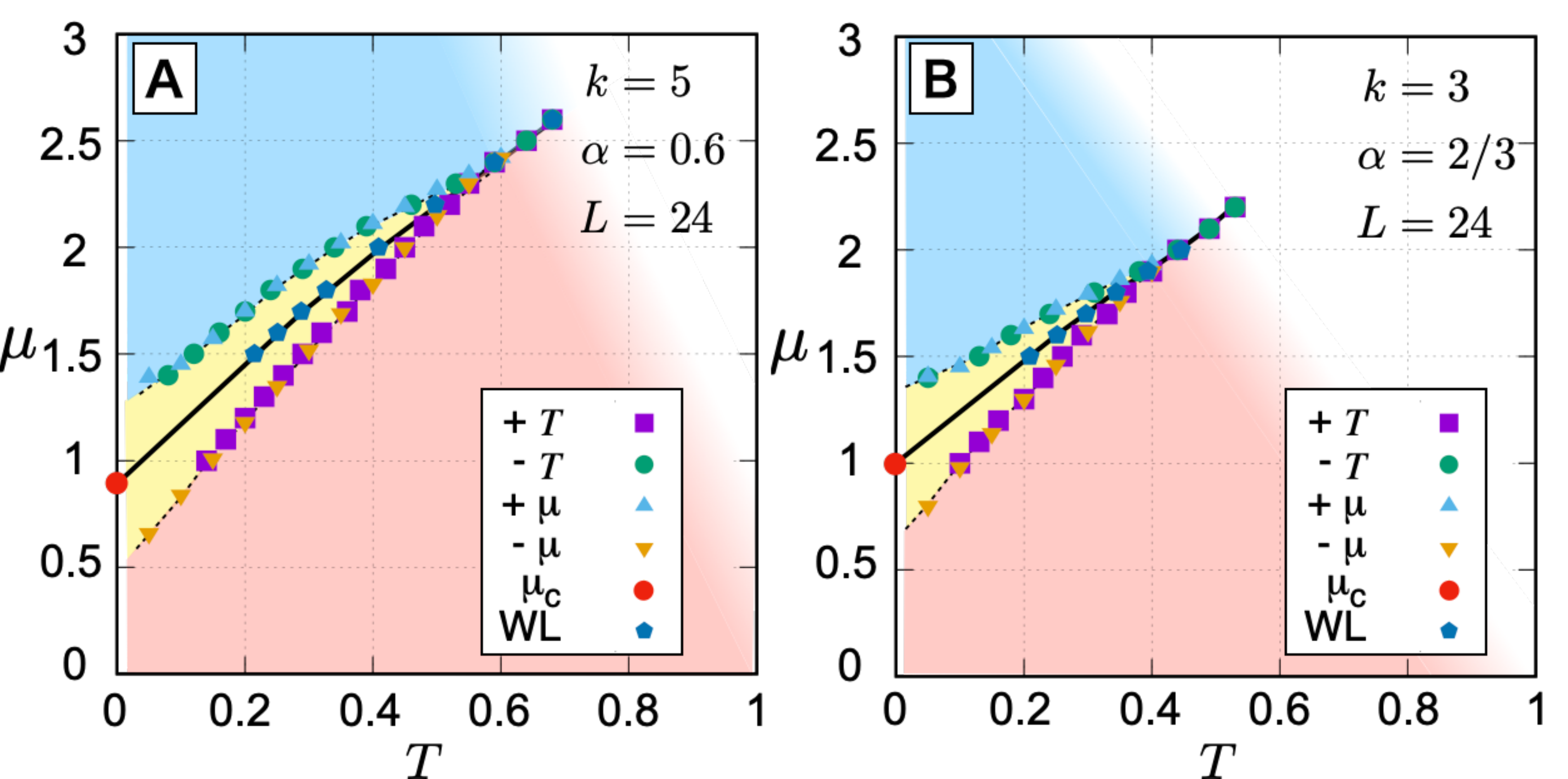}
\caption{
Phase diagram.
({\it A}) Phase diagram of the model with $k=5$ and $\alpha=0.6$.
The solid curve represents the equilibrium phase boundary between the adsorbed and desorbed phases, determined from the specific-heat peaks obtained by the WL method. The transition point $\mu_{\rm c}=3k\alpha^2/(1+k)$ at $T=0$ is determined analytically by comparing the minimum energy of $V_1$ and $V_1+V_2-\mu$. The boundaries between different colors are determined from the specific-heat peaks obtained by quasi-equilibrium protocols; heating $(+T)$, cooling $(-T)$, increasing chemical potential $(+\mu)$, and decreasing chemical potential $(-\mu)$.
({\it B}) Phase diagram for $k=3$ and $\alpha=2/3$. The equilibrium swelling ratio is the same as {\it A}, but the hysteresis shrinks.
The system size $L=24$.
}
\label{fig:phase}
\end{figure}

\subsection*{Thermodynamics}
\blue{We mainly present the results of adsorption expansion and hardening case, i.e., $k>0$ and $k\alpha>0$.}
First, we examine the effect of adsorption hardening on the phase diagram. Hereafter the temperature $T$ and chemical potential $\mu$ are expressed in units of \blue{$k_0\ell_0^2/k_{\rm B}$} and $k_0\ell_0^2$, where $k_{\rm B}$ is Boltzmann constant. The $(T, \mu)$ phase diagram is presented in Fig.~\ref{fig:phase}, where $(k,\alpha)=(5,0.6)$ in Fig.~\ref{fig:phase}{\it A} and $(3,2/3)$ in Fig.~\ref{fig:phase}{\it B}. The equilibrium swelling ratio is the same when all sites adsorb guest particles, but the adsorbed state becomes more rigid in Fig.~\ref{fig:phase}{\it A}. Adsorbed and desorbed phases are realized in the red and blue regions, respectively. They are independent of the simulation protocols, whereas hysteretic behavior is observed in the yellow region. The equilibrium phase boundary obtained by the WL simulations crosses the middle of the hysteretic region. Below (above) the equilibrium phase boundary in the yellow region, the adsorbed (desorbed) phase is not thermodynamically stable. Alternatively, they become metastable that are stable against infinitesimal fluctuation. The critical points are $(T,\mu)\simeq(0.7,2.6)$ in Fig.~\ref{fig:phase}{\it A} and $(T,\mu)\simeq(0.53,2.2)$ in Fig.~S\ref{fig:phase}{\it B} (see \blue{{\it Appendix}, Fig.~S\ref{fig:supple_diagram}} for the cases of other $k$ and $\alpha$). The critical temperature increases and the hysteretic region becomes broader for larger $k$, implying that the phase transition between the adsorbed and desorbed phases becomes strongly first-order when adsorption hardening occurs. Note that the system size slightly affects the transition temperature when $L\ge 24$ adopted in this study, while strong size dependencies are observed for smaller system sizes~\cite{sakata2013shape} (see \blue{{\it Appendix}, Fig.~S\ref{fig:supple_size}}). Thus, from a microscopic (macroscopic) perspective, the hysteretic response can be controlled by changing the magnitude of the host-guest interaction (differences in the elastic moduli between the adsorbed and desorbed phases). We focus on strong elastic heterogeneity $k=5$ and $\alpha=0.6$ because the role of elastic heterogeneity becomes noticeable.

\begin{figure}[t]
\centering
\includegraphics[width=85mm]{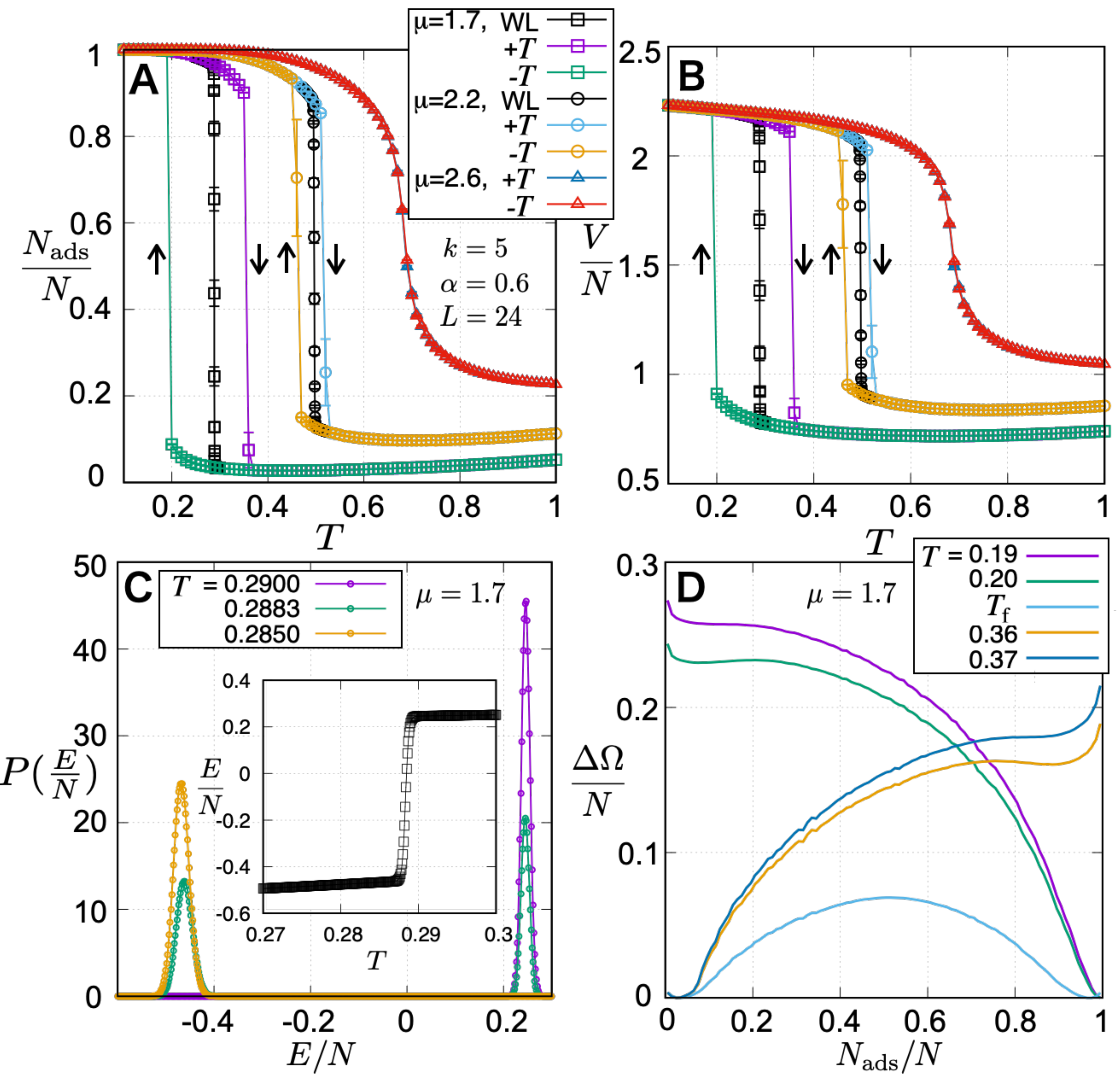}
\caption{
Thermodynamics of the adsorption/desorption transition.
({\it A} and {\it B}) Temperature dependencies of the adsorption fraction and the volume per site obtained by the WL method, heating $(+T)$ and cooling $(-T)$ protocols for $k=5$, $\alpha=0.6$, and $L=24$.
({\it C}) Probability distribution of the total energy around the first-order transition point $T_{\rm f}=0.2883$, $\mu=1.7$. The inset shows the temperature dependencies of thermally averaged energy.
({\it D}) The osmotic grand potential $\Delta \Omega(N_{\rm ads}/N)=\Omega(N_{\rm ads}/N) - \Omega_{\rm min}$ at $T=0.19$, $0.20$, $T_{\rm f}$, $0.36$, $0.37$ for $\mu=1.7$, where $\Omega_{\rm min}$ is the minimum value at each $T$ (see {\it Materials and Methods} for the definitions of $P(E/N)$ and $\Delta\Omega$). The error bars in {\it A} and {\it B} represent the standard error.
}
\label{fig:physical}
\end{figure}

\begin{figure}[t]
\centering
\includegraphics[width=85mm]{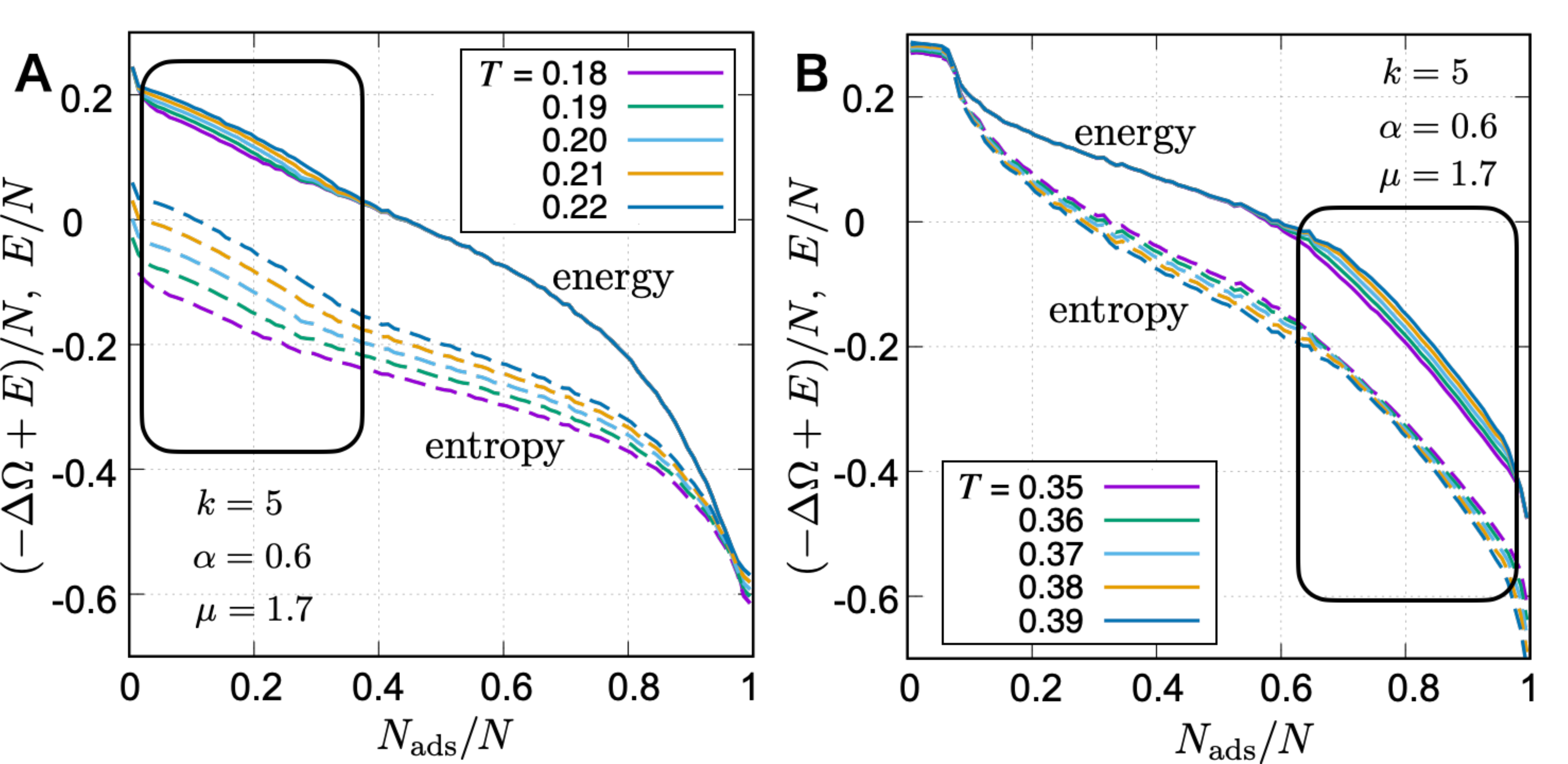}
\caption{
Energetic and entropic contributions to the osmotic grand potential.
We decompose the osmotic grand-potential landscape $\Delta\Omega(N_{\rm ads}/N)/N$ into energetic $E/N$ (the solid curves) and entropic $TS/N=(-\Delta\Omega+E)/N$ (the broken curves) contributions.
({\it A}) Energetic and entropic contributions close to the adsorption transition point in the cooling process.
The inside of the frame represents the metastable desorbed state, which becomes unstable for $T<0.20$.
({\it B}) Energetic and entropic contributions close to the desorption transition point in the heating process.
The inside of the frame represents the metastable adsorbed state, which becomes unstable for $T>0.36$.
$\mu=1.7$, $k=5$, and $\alpha=0.6$
}
\label{fig:supple_landscape}
\end{figure}

Next, we examine the thermal stability of the adsorbed and desorbed phases. In the standard MC calculations, the transition between the adsorbed and desorbed states exhibits large hysteresis. The temperature and chemical potential dependencies of the adsorption fraction $N_{\rm ads}/N$ ($N_{\rm ads}$ is the number of the adsorbed particles) and volume per site $V/N$ are presented in Fig.~\ref{fig:physical}{\it A} and {\it B}, respectively. Discontinuous changes are confirmed around the transition temperatures for $\mu=1.7$ and $2.2$. As $\mu$ increases, the discontinuity weakens, and eventually vanishes above the critical point. By increasing the temperature above the transition temperature, the adsorption fraction gradually increases to $0.5$ because the entropy with respect to the distribution of the guest molecules maximizes. The system volume behaves in the same manner as $N_{\rm ads}$, as shown in Fig.~\ref{fig:physical}{\it B}. We also note that MC simulations with varying chemical potential protocols exhibit a similar tendency (see \blue{{\it Appendix}, Fig.~S\ref{fig:mu}}).
\blue{We also perform MC simulations for the other combinations of the signs of $k\alpha$ and $k$, i.e., adsorption-induced lattice expansion and softening, contraction and softening, and contraction and hardening cases. Although the slope of the phase boundaries is different for these cases, hysteretic behavior is observed in all cases (see \blue{{\it Appendix}, Fig.~S\ref{fig:supple_diagram_kalpha}}).}

To examine the nature of the first-order transition more deeply, we also perform WL calculations in the osmotic ensemble~\cite{bousquet2012free} (see {\it Materials and Methods} for details). The computed probability distribution of energy $E$ and the thermal average of the total energy are presented in Fig.~\ref{fig:physical}{\it C}. From the inset, we obtain the equilibrium transition temperature $T_{\rm f}=0.2883$ at which the isobaric specific heat $C_P$ is maximized. At $T=T_{\rm f}$, the probability distribution exhibits two peaks at $E/N=-0.45$ and $0.25$, corresponding to the adsorbed and desorbed states, respectively. The probability distribution of metastable states decreases dramatically when the temperature changes slightly to $T=0.290$ and $T=0.285$. Nevertheless, they remain finite over a broad temperature range, which is consistent with the appearance of the hysteresis loop. Fig.~\ref{fig:physical}{\it D} shows the osmotic grand-potential landscape with respect to the adsorption fraction at $\mu=1.7$. At $T=T_{\rm f}$, the landscape exhibits a doubly-degenerate structure, indicating the coexistence of adsorbed and desorbed phases in equilibrium. The degeneracy lifts by temperature change, and eventually, the metastability of the desorbed (adsorbed) phase vanishes below $T=0.20$ (above $T=0.36$). 
These spinodal points coincide with the transition temperatures in the standard MC simulations within our simulation time (see {\it Materials and Methods} for the simulation protocols), implying that the nucleation is strongly suppressed in metastable states. 
\blue{When we take longer simulation time, the transition temperature departs from the spinodal point (see \blue{{\it Appendix}, Fig.~S\ref{fig:supple_size}}).}
By decomposing the osmotic grand potential into energetic and entropic contributions, a conspicuous asymmetry between the adsorption transition and desorption transition is observed, as shown in fig.~\ref{fig:supple_landscape}.
The entropy decrease is much larger than the energy decrease in the metastable desorbed state in the cooling process, whereas the energy increase is more significant than the entropy decrease in the metastable adsorbed state. This suggests that the entropic (energetic) contribution is responsible for the adsorption (desorption) transition.
In a subsequent section, we discuss that the robustness of the hysteresis and asymmetric transition behavior result from elastic heterogeneity.

The multicanonical MC simulation result explains the slope of the equilibrium phase boundary between the adsorbed and desorbed phases presented in Fig.~\ref{fig:phase}. At $(\mu,T)=(1.7,T_{\rm f})$, the osmotic grand potential difference vanishes, and the energy difference per particle is 0.7, leading to the entropy difference $\Delta S/N=0.7/T_{\rm f}$. The Clausius-Clapeyron equation in $T-\mu$ plane (Fig.~\ref{fig:phase}) at $\mu=1.7$ reads $(d\mu/dT)_{\rm cx}=-\Delta S/\Delta N_{\rm ads}=(0.7/0.2883)/0.9 \sim 2.7$, where the subscript cx denotes differentiation along the coexistence curve (see {\it Materials and Methods}, Eq.1). 
\blue{Although the slope is always positive when lattice expansion and hardening are induced by adsorption, the situation changes in other cases (see \blue{{\it Appendix}, Fig.~S\ref{fig:supple_diagram_kalpha}}).
The slope can be positive and negative by varying $k$ and $\alpha$, implying that the sign of the entropy difference $\Delta S$ depends on systems.}

\begin{figure}[t]
\centering
\includegraphics[width=85mm]{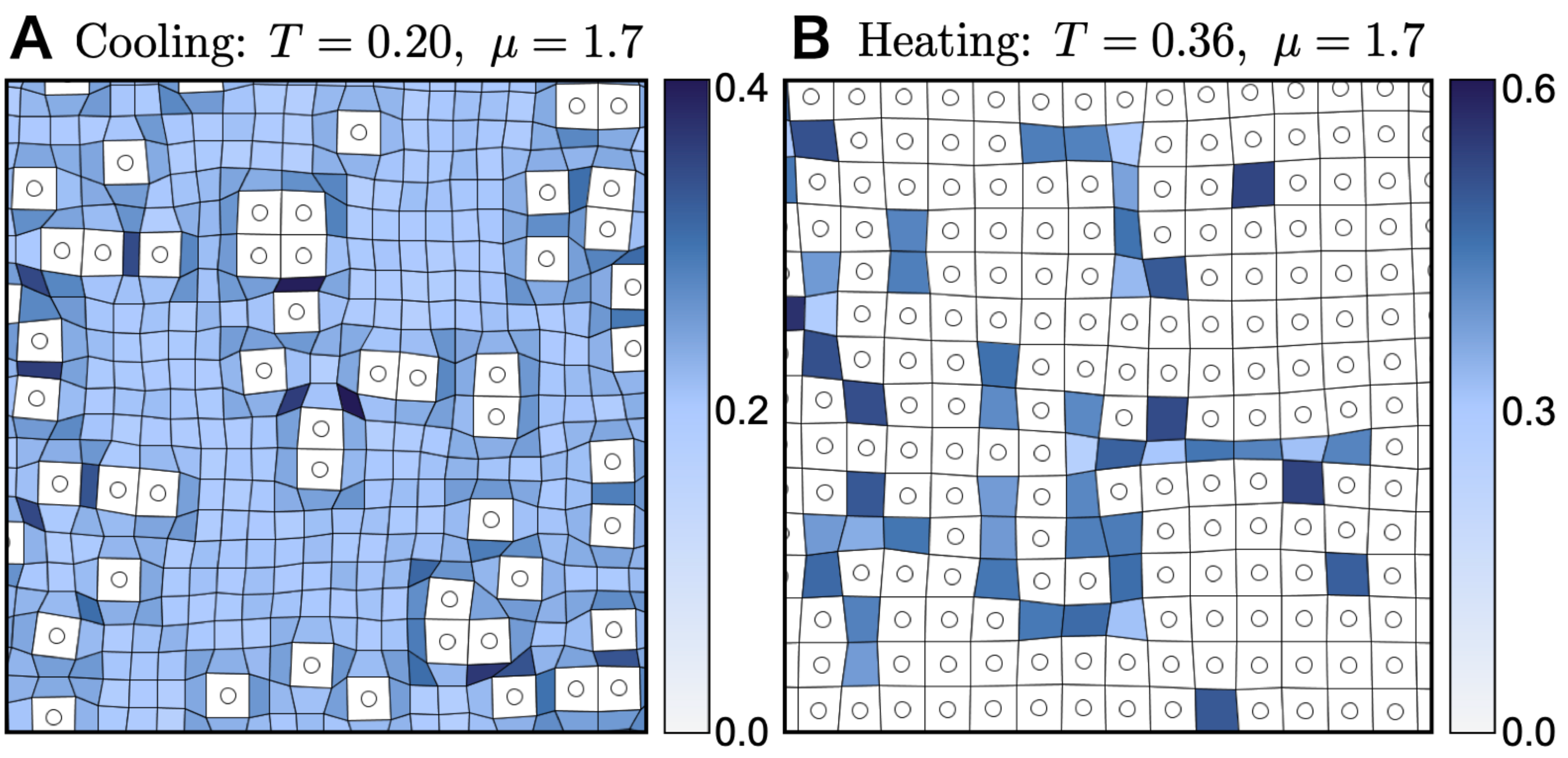}
\caption{
Heterogeneous distribution of the adsorbates and the elastic energy.
({\it A}) Snapshot of an equilibrated desorbed state obtained at $(T,\mu)=(0.20,1.7)$ slightly above the adsorption transition temperature in the cooling process.
({\it B}) an equilibrated adsorbed state obtained at $(T,\mu)=(0.36,1.7)$ slightly below the desorption transition temperature in the heating process.
Circles denote the adsorbed sites. The color in the desorbed sites represents their energy. Short-time averaging over 1000 MCSs with fixed adsorbate distribution is performed to obtain the average lattice distortion and elastic energy. $k=5$, $\alpha=0.6$, and $L=96$.
}
\label{fig:snapshot}
\end{figure}

\subsection*{Elastic heterogeneity determines the distribution of adsorbates}
Now let us investigate the origin of the robustness of the hysteresis and asymmetric energy-entropy contributions by examining the role of elastic heterogeneity. Fig.~\ref{fig:snapshot}{\it A} shows a snapshot at $(T,\mu)=(0.20,1.7)$ in the metastable desorbed state, slightly above the adsorption transition temperature. At this point, the adsorbed fraction is $N_{\rm ads}/N=0.08$. Since the adsorption of the guest particles induces lattice expansion, a lattice mismatch between the adsorbed and desorbed sites emerges. Indeed, the cell volume of the adsorbed (desorbed) sites becomes $2.02\pm 0.04$ ($0.80\pm 0.08$), while their natural sizes are $(1+k\alpha/(1+k))^2=2.25$ and $1$, respectively. The lattice mismatch relaxes by the deformation of the harder adsorbed sites ($\sim 10\%$ contraction) and softer desorbed sites ($\sim 20\%$ contraction), with the latter undergoing a larger deformation to reduce the overall elastic energy. The characteristic adsorbate distribution arises from elasticity-mediated long-range interactions between them. The interaction between the two adsorption sites is strongly anisotropic. They are repulsive when diagonally adjacent, whereas they are attractive when horizontally and vertically adjacent (see \blue{{\it Appendix}, Fig.~S\ref{fig:supple_effective}{\it A}}). This anisotropy results in energetic frustration between three adsorbates. In this case, the trimer formation is energetically unfavorable. Alternatively, a configuration in which a dimer and a monomer are separated by one desorbed site is energetically stable (see \blue{{\it Appendix}, Fig.~S\ref{fig:supple_effective}{\it B}}). The situation becomes more complicated in the case of clusters larger than tetramer. If the cluster's shape is not isotropic, the energy becomes higher than that of dispersed monomers (see \blue{{\it Appendix}, Fig.~S\ref{fig:domain_size}{\it A}}). In phase ordering without elasticity, cluster aggregation reduces the energy because the surface area decreases~\cite{binder1987theory,Onukibook}. This cannot be adopted in our case since the deviation of the domain shape from the isotropic form is inhibited. This feature is reminiscent of the inclusion problem in phase-separating alloys, where it was shown by Eshelby that hard domains embedded in a soft matrix prefer isotropic shapes~\cite{eshelby1957determination,Khachaturyan,Onukibook}. Since the adsorbed domains are more rigid due to the presence of $V_2$, Eshelby's argument holds in our system (see \blue{{\it Appendix}, Fig.~S\ref{fig:nads_fix}{\it A}}). The isotropic domains appear selectively, reducing the number of stable domain morphology, hence reducing the entropy. Thus the growth of the adsorbed domains is hindered entropically.
The inter-domain interaction is also constrained due to the energetic frustration arising from the long-range nature of the elastic interaction. Thus, the entropy of stable domain dispersion states also decreases as the temperature is lowered, leading to the spinodal instability.

In the heating process, on the other hand, the morphology of the desorbed domains in the adsorbed state differs from that of the adsorbed domains in the desorbed state. A snapshot at $(T,\mu)=(0.36,1.7)$ in the metastable adsorbed state, slightly below the desorption transition temperature, is displayed in Fig.~\ref{fig:snapshot}{\it B}. At this point, the desorbed fraction is $1-N_{\rm ads}/N=0.12$. Contrary to the cooling process, the desorbed domains preferentially form narrow channels. This can be understood by examining the interaction between desorbed sites. Similar to the case of adsorbed sites, two desorbed sites prefer to aggregate to reduce the elastic energy (see \blue{{\it Appendix}, Fig.~S\ref{fig:supple_effective}{\it C}}). For three desorbed sites, they also prefer to aggregate; however, a notable difference arises. The formation of a straight channel is energetically favored (Fig.~\ref{fig:supple_effective}{\it D}). This tendency also holds for larger domains, while other configurations are also more favorable than the vacancies-separated state (see \blue{{\it Appendix}, Fig.~S\ref{fig:domain_size}{\it B}}). This is because the desorbed domains are softer than the surrounding adsorbed region. It was also shown by Eshelby that a soft inclusion in a harder matrix changes its shape to a flat domain to reduce the elastic energy~\cite{eshelby1957determination,Khachaturyan,Onukibook} (see \blue{{\it Appendix}, Fig.~S\ref{fig:nads_fix}{\it B}}). The flattening of the domain morphology increases the interfacial area compared with isotropic domains, resulting in an increase in the interfacial energy. The stability of the desorbed domain is reduced energetically, preventing domain growth. Thus, we may conclude that elastic heterogeneity controls the domain growth and morphology asymmetrically, stabilizing the metastable desorbed (adsorbed) states against nucleation entropically (energetically) in the cooling (heating) processes.

\begin{figure}[t]
\centering
\includegraphics[width=85mm]{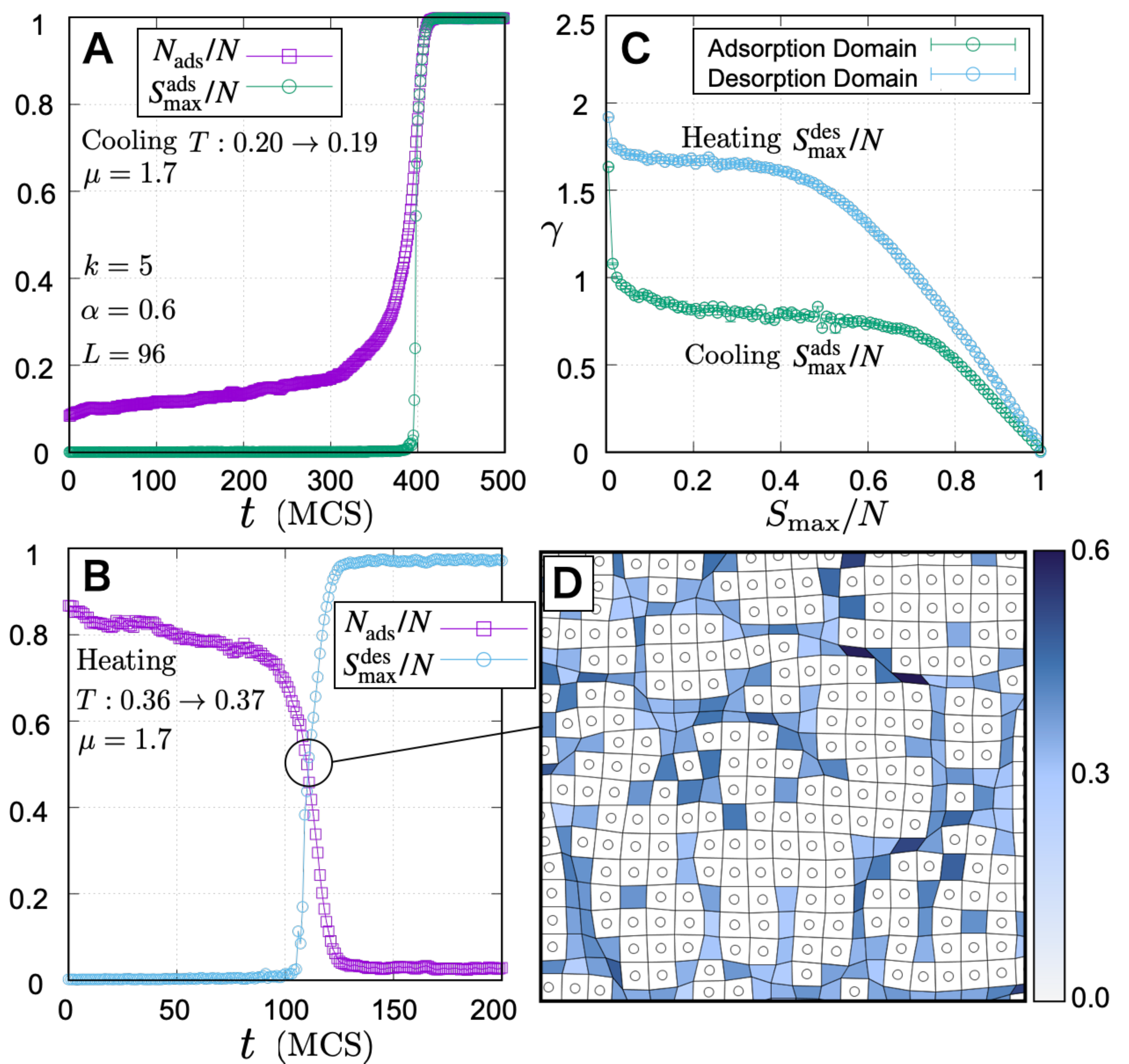}
\caption{
Kinetics of the transition between the adsorbed and desorbed phases.
({\it A} and {\it B}) The time evolution of the adsorption fraction $N_{\rm ads}/N$ and the growing maximum domain sizes $S_{\rm max}/N$ at the spinodal points in the cooling process ({\it A}) and heating process ({\it B}).
Initially, metastable desorbed state at $T=0.2$ ({\it A}) and metastable adsorbed state at $T=0.36$ ({\it B}) are equilibrated. Then, the transition kinetics into adsorbed ({\it A}) and desorbed ({\it B}) states by changing temperature to $T=0.19$ ({\it A}) and $0.37$ ({\it B}) are examined.
({\it C}) The boundary length per site $\gamma$ of the growing maximum size domain with respect to the domain size $S_{\rm max}$ averaged over 150 independent runs for both cooling and heating processes.
({\it D}) Typical snapshot in the transition process. The notation is the same as Fig.~\ref{fig:snapshot}.
$k=5$, $\alpha=0.6$, $\mu=1.7$, and $L=96$.
}
\label{fig:domain}
\end{figure}

The role of elastic heterogeneity becomes conspicuous in transformation kinetics. We examine the MC kinetics of the time evolution of adsorption (desorption) during the cooling (heating) process across the transition temperature. In Fig.~\ref{fig:domain}{\it A} and B, the time evolution of the adsorption and desorption fraction and the size of the largest domain are displayed. The largest domain grows steeply around $t=400$ MC steps (MCS) in Fig.~\ref{fig:domain}{\it A}, and $t=100$ MCS in Fig.~\ref{fig:domain}{\it B}, respectively, though the adsorption density gradually changes even before the steep growth. This result implies that domain growth occurs due to the abrupt coalescence of small domains. The boundary length per site $\gamma$ of the largest domain increases when the domain exhibits an anisotropic shape. Fig.~\ref{fig:domain}{\it C} indicates that $\gamma$ becomes larger for the desorbed domain in the heating process than for the adsorbed domain in the cooling process, characterizing the anisotropic (isotropic) domain growth in the heating (cooling) process. The anisotropy of the growing desorbed domains at the transient state is illustrated in Fig.~\ref{fig:domain}{\it D}, where the desorbed sites form percolated domains (see \blue{{\it Appendix}, Movie~S1} for the spatial pattern evolution during the transformation). In the cooling process, on the other hand, dispersed domains grow isotropically, which eventually coalesce to form a single large adsorbed domain (see \blue{{\it Appendix}, Movie~S2}). Thus, elastic heterogeneity leads to the asymmetric growth kinetics of the adsorbed and desorbed domains.

\blue{Such intermediate adsorption states can also be obtained by simulations under the fixed adsorption fraction. It is confirmed that harder/softer domains form isotropic/anisotropic shapes in four cases: lattice expansion and hardening, lattice expansion and softening, lattice contraction and softening, and lattice contraction and hardening conditions (see \blue{{\it Appendix}, Fig.~S\ref{fig:nads_fix}}), indicating that Eshelby's argument generally holds.}

\section*{Discussion}

\begin{table*}[t]
\centering
\caption{Mechanical parameters of three soft porous crystals estimated from experiments and density functional theory: $B$ and $V_{\rm cell}$ represent the bulk modulus and unit cell volume adopted from literature, where np (cp) and lp (op) represent the narrow (closed) pore and large (open) pores, respectively. $k$ and $\alpha$ regarding structural transformations are calculated from $B$ and $V_{\rm cell}$. $E_{\rm eff}$ is the energy scale of the effective guest-guest interactions due to elastic heterogeneity, estimated from $B$, $V_{\rm cell}$, $k$, and $\alpha$. We apply $B=4.29$ for bulk modulus in the np phase of MIL-53(Cr) to evaluate $k,~\alpha$ and $E_{\rm eff}$.}
\label{tab:material}
\begin{threeparttable}
\begin{tabular}{|c|c|c|c|c|c|c|c|c|c|c|c|c|}
\hline
~ & \multicolumn{2}{|c|}{np (cp)} & \multicolumn{2}{|c|}{lp (op)} & \multicolumn{3}{|c|}{np (cp) $\to$ lp (op)} & \multicolumn{3}{|c|}{lp (op) $\to$ np (cp)} \\
\hline
material & $B$ (GPa) & $V_{\rm cell}$ (\AA$^3$) & $B$ (GPa) & $V_{\rm cell}$ (\AA$^3$) & $k$ & $\alpha$ & $E_{\rm eff}$ ($10^{-21}$J) & $k$ & $\alpha$ & $E_{\rm eff}$ ($10^{-21}$J) \\
\hline
MIL-53(Al) & 10.7\tnote{a} & 939.92\tnote{a} & 0.35\tnote{b} & 1423.8\tnote{b} & $-0.967$ & $-0.0051$ & 0.90 & 30.6 & $-0.155$ & 2.10 \\
\hline
MIL-53(Cr) & 4.29\tnote{a},~10\tnote{c} & 987.2\tnote{a} & 2\tnote{c} & 1486\tnote{d} & $-0.53$ & $-0.13$ & 24.1 & 1.15 & $-0.27$ & 25.0 \\
\hline
DUT-49\tnote{d} & 5.6 & 46070 & 7.5 & 96047 & 0.34 & 1.09 & 8771.7 & $-0.25$ & 0.83 & 18008.8 \\
\hline
\end{tabular}
\begin{tablenotes}
\item[a] Taken from~\cite{yot2016impact}. $B$ and $V_{\rm cell}$ are determined by fitting experimental data by using the Murnaghan equation of state.
\item[b] Taken from~\cite{yot2014metal}. $B$ and $V_{\rm cell}$ are measured at room temperature.
\item[c] Taken from~\cite{neimark2011structural}. $B$ is extracted from mercury intrusion--extrusion and $V_{\rm cell}$ is determined from crystallography. 
\item[d] Taken from~\cite{ying2021effect}. $B$ and $V_{\rm cell}$ are obtained by density functional theory at $T=0$ (K).
\end{tablenotes}
\end{threeparttable}
\end{table*}

\blue{In our model, an effective guest-guest interaction due to elastic heterogeneity results in an adsorption/desorption transition with hysteresis. In real SPCs, however, direct guest-guest interactions such as van der Waals and electrostatic forces also exist, which are not incorporated into our model. From thermodynamic arguments for general adsorption problems, it has been shown that hysteretic adsorption/desorption can occur when guest-guest intermolecular interactions are strong~\cite{oliveira1978lattice,casal2002phase,borkovec2008electrostatic}. Therefore, in this section, we provide an order estimation of the effective interactions arising from elastic heterogeneity in real SPCs and discuss the relevance of the elasticity-mediated interactions on hysteretic adsorption/desorption.}

\blue{The effective interaction between guest particles due to elastic heterogeneity depends on the model parameters $k$ and $\alpha$ and energy unit $k_0\ell_0^2$, where $k_0$ and $\ell_0$ are the spring constant and lattice constant of the host matrix in the desorbed phase. By taking the second derivative in the Hamiltonian, the bulk modulus in our model reads $k_0$ and $(1+k)k_0$ in the fully desorbed and adsorbed phase, respectively. The model parameters can be estimated from the ratio of the bulk modulus between the desorbed and adsorbed phases and the ratio of the lattice constants reported in the literature (Table~1). Since the bulk modulus of the fully desorbed phase in our model is $B^{\rm des}=k_0$, the energy unit $k_0\ell_0^2$ becomes the bulk modulus $B^{\rm des}$ multiplied by the unit cell volume $V_{\rm cell}^{\rm des}$ in the desorbed phase. Thus, the energy scale of the effective interaction mediated by elasticity is given by $E_{\rm eff} = U_{\rm eff}(k,\alpha) \times B^{\rm des}V_{\rm cell}^{\rm des}$, where $U_{\rm eff}(k,\alpha)$ is a numerical factor depending on $k$ and $\alpha$. We envisage $U_{\rm eff}(k,\alpha)$ as the elastic energy difference (in the unit of $k_0\ell_0^2$) between an adsorption dimer and two completely isolated adsorption sites. $U_{\rm eff}(k,\alpha)$ is typically of the order of $10^{-2}$, proportional to $\alpha^2$ and behaves non-monotonically with respect to $k$ (see \blue{{\it Appendix}, Fig.~S\ref{fig:supple_Ueff}}).}

\blue{Now, we consider MIL-53($M={\rm Al, Cr}$) and DUT-49, whose lattice parameters and bulk modulus have been reported~\cite{coudert2016computational,yot2016impact,ferey2009large,krause2016pressure,yot2014metal,ying2021effect,neimark2011structural}. These substances exhibit adsorption/desorption transition with structural transformation by gas loading. The small and large pore phases are termed the narrow pore (np) and large pore (lp) for MIL-53 and the closed pore (cp) and open pore (op) for DUT-49. The fully desorbed phase is the lp and op phase in MIL-53 and DUT-49, respectively. As increasing the loading, a structural phase transition to the np and cp states occur. By further increasing the loading, they return to the lp and op states. That is, the lp and op (np and cp) phases are the desorbed (adsorbed) phases in the former, whereas adsorbed (desorbed) phases in the latter. Therefore, we estimate $E_{\rm eff}$ both in np (cp) to lp (op) phases and lp (op) to np (cp) phases, as shown in Table~1. From the table, $E_{\rm eff}$ is the smallest in MIL-53(Al) because $|\alpha|$ is the smallest. Even in this substance, the elasticity-mediated interaction $E_{\rm eff}$ is comparable to the van der Waals force, which is the order of $10^{-21}$J~\cite{Israelachvili}. MIL-53(Cr) and DUT-49 have much larger $E_{\rm eff}$. Thus, the effect of elastic heterogeneity should be prominent, which may lead to large hysteresis, though electrostatic guest-guest interactions can be stronger when adsorbed molecules are polarized (e.g., hydrogen-bond interaction is $\sim 30~(10^{-21}{\rm J})$ for water molecules~\cite{Israelachvili}). In contrast, some other SPCs, such as ZIF-8~\cite{zhang2013sorption}, exhibit small structural deformation during the adsorption transition, resulting in very small $|\alpha|$. Elastic heterogeneity would be less relevant in these substances.}

\blue{
Finally, we discuss the dependency of our result on model parameters and dimensionality. In this study, we assume that the NN and NNN interactions have the same spring constant $k_0$. In real systems, their ratio can deviate from unity. However, the qualitative feature of the phase diagram remains unchanged when the NNN interaction is weaker than NN, though the hysteretic region shrinks as the NNN interaction becomes weak (see \blue{{\it Appendix}, Fig.~S\ref{fig:supple_diagram_NNN}}).
In three-dimensional systems, furthermore, we confirm that similar hysteresis is observed when our model is extended to a three-dimensional simple cubic lattice (see \blue{{\it Appendix}, Fig.~S\ref{fig:supple_3d}}). Compared to the two-dimensional model, a broad hysteresis is observed for smaller $k$ and $\alpha$ in the three-dimensional model. According to the statistical mechanical argument, this may arise from the suppression of thermal fluctuations as the number of interacting pairs increases in higher dimensions. Thus, the effect of elastic heterogeneity would be pronounced in real three-dimensional SPCs.}

\section*{Summary}
In conclusion, we elucidated the asymmetric role of elastic heterogeneity ---the difference in the lattice constant and mechanical rigidity--- on the adsorption-desorption transition. 
The result that the domain shape depends on the difference in the elastic stiffness suggests promising applications. If adsorption hardening occurs, the domain favors a compact isotropic shape, implying strong confinement and dispersion of the adsorbates inside host crystals. This can be applied to gas storage without leakage, enzyme immobilization, and supports inclusions such as nanoparticles without sintering.
If adsorption softening occurs, on the other hand, the domain prefers a flattened shape, increasing the surface area. A large interfacial area between the adsorbed and desorbed sites can be utilized to enhance catalytic reactions. Moreover, enhancing the intrusion and transport of the adsorbates would be possible since they prefer to form percolated domains, though further molecular dynamics investigation should be conducted to reveal the transport kinetics. Thus, controlling elastic heterogeneity can provide a guideline for practical applications.
In this study, we focused on isotropic swelling, preserving the crystalline nature of the substances under gas adsorption and desorption. Incorporating anisotropic deformation and elastic moduli~\cite{Khachaturyan,Onukibook,ortiz2012anisotropic} and disorder~\cite{furukawa2015heterogeneity,bennett2021changing,kartha1991spin} to examine the impact of elastic heterogeneity should be a further research direction.
Finally, from the theoretical viewpoint, our model exhibits phase transitions with respect to a non-conserved order parameter (the adsorption fraction in this study) coupled with crystal elasticity. This feature is not limited to MOFs but can also be applied to other elastic systems such as magnetic skyrmions, where elastic softening occurs around the topological phase transition~\cite{nii2014elastic}. We believe our method provides a route to examine the role of the elastic heterogeneity on phase transitions in these systems.

\section*{Method}

\subsection*{Average swelling ratio and volume}
Because the average size of the simulation cell varies with the gas adsorption/desorption, the system volume $V$ also varies during the simulations. Therefore, we impose periodic boundary conditions in $x$ and $y$ directions such that the system becomes periodic under the translation of $aL$. Here, $a=\sqrt{V/V_0}$ is the average swelling ratio, and $V_0=N\ell_0^2$ is the reference system volume.
\subsection*{Unit Monte Carlo step}
The unit Monte Carlo (MC) step consists of one Metropolis sweep for the adsorption/desorption of guest particles $\{\sigma_\square\}$, $L$ iterations of Metropolis sweeps for the lattice sites $\{\bm{r}_i\}$, and $L$ iterations of Metropolis updates for the affine displacement of the system to change $a=\sqrt{V/V_0}$. Thus, the unit MC step consists of $N+L\times N+L$ Metropolis updates. The updates of $\bm{r}_i$ and $a$ are restricted to $|\Delta \bm{r}_i|<0.1$ and $|\Delta a|<0.01$, respectively. During the updates of the lattice sites $\{\bm{r}_i\}$, displacements causing bond intersects are prohibited to preserve the square lattice configuration without folding.

\subsection*{Standard Monte Carlo simulation}
To capture the hysteretic behavior of the adsorption-desorption transitions, we perform a standard MC simulation, where the temperature $T$ and chemical potential $\mu$ are varied quasistatically. In the standard MC simulations, we iterate $10^4$ MC steps for equilibration, and subsequently, $10^4$ MC steps for obtaining a thermal average at each $T$ and $\mu_{\rm ads}$. Subsequently, we incrementally change the temperature and chemical potential $\Delta T=0.01$ and $\Delta \mu=0.01$, respectively. In the cooling simulations, we prepare the initial configurations at a sufficiently high temperature $T=1$. In the heating and decreasing $\mu$ simulations, we construct the initial configuration such that all the plaquettes are occupied by the guest particle, and the elastic energy is minimized. In the increasing $\mu$ simulations, we prepare the initial configuration such that all plaquettes are not occupied and the elastic energy is minimized. We perform five independent runs for each protocol and evaluate statistical errors.

\subsection*{Multicanonical Monte Carlo simulation}
To examine the equilibrium phase transitions shown in Figs.~\ref{fig:phase} and \ref{fig:physical}, the equilibrium probability distribution of the energy $E$ and the adsorption density $n=N_{\rm ads}/N$ as a function of $\beta=1/k_{\rm B}T$ and $\nu=\mu_{\rm ads}/T$ must be obtained (we fix the pressure $P=0$ as described in the main text).
For this purpose, we adapt multicanonical MC simulations employing the Wang-Landau (WL) method~\cite{wang2001efficient,Landau-Binder,bousquet2012free}. In the WL method, we calculate the probability distribution function of energy $E$. Let us consider the probability $P(\bm x)$ of the microscopic state $\bm x$. If $P(\bm x)$ is uniform, every microscopic state is sampled with equal probability. However, it is computationally inefficient to calculate the probability distribution function of $E$ by uniform $P(\bm x)$.
Alternatively, in the WL method, $P(\bm x)$ is proportional to $e^{-g(E)}$, where $g(E)$ is a weight function. By performing the preliminary run described below, we obtain $g(E)\cong S(E)+{\rm const.}$, where $S(E)$ is the entropy. Hence, the energy histogram $H(E)=\sum_{\{\bm x | {\cal H}(\bm x)=E\}}P(\bm x)$ becomes uniform with respect to the energy according to statistical mechanics~\cite{Landau5}, where ${\cal H}$ is the Hamiltonian defined in Eq.~\ref{eq:hamiltonian}. This weighted probability enables us to calculate the probability distribution function of energy efficiently. The obtained $P(\bm x)$ is utilized to calculate the equilibrium probability distribution of ($E$, $n$) and the thermal average of the physical quantities at arbitrary temperatures.

The preliminary run in this study comprises seven steps.
\begin{enumerate}
\item Divide the energy range $E_{\rm GS}/N + (1-E_{\rm GS}/N)/200 \le E/N \le 1$ into 1000 bins, where $E_{\rm GS}$ is the ground state energy. For each bin, the histogram $H_i$ and $g_i$ are initialized to $0$ ($i$ is the index of a bin). We also set the increment $\Delta g$ to be $1$.
\item The unit MC step described above is performed. At each Metropolis update in the unit MC step, the energy range to which the energy of the trial state belongs is noted as $i$. The acceptance rate of the trial state reads $\min\{1,e^{-(g_i-g_j)}\}$, where the subscript $j$ stands for the state before the update. The trial state with its energy being outside the energy range defined in the step $1$ is also rejected. If accepted, we update $H_i \rightarrow H_i+1$ and $g_i \rightarrow g_i+\Delta g$; otherwise, we update $H_j \rightarrow H_j+1$ and $g_j \rightarrow g_j+\Delta g$.
\item Continue the update until $\min_i H_i\ge (4/5)\sum_i H_i /1000$ is satisfied.
\item $H_i$ for all $i$ is set to $0$, and $\Delta g$ is divided by $2$.
\item Steps 2-4 are repeated until $\Delta g$ becomes less than $10^{-6}$.
\item Sample the energy histogram $H_i$ using $g_i$ for all $i$ with $5 \times 10^6$ MC steps.
\item Correct the weight function as $\tilde{g}_i = g_i + \log H_i$.
\end{enumerate}
By performing the preliminary run, the histogram $H$ becomes almost uniform; hence, $\tilde{g}_i \cong S(E_i) + {\rm const.}$. 

To compute the equilibrium probability distribution of ($E$, $n$), we divide the energy range $E_{\rm GS}/N + (1-E_{\rm GS}/N)/200 \le E/N \le 1$ into 1000 bins and the adsorption fraction $0\le n\le 1$ into 100 bins.
In each bin, the averages of physical quantities $\bar{A}_{E_i, n_j}$ ($A$ is the energy, adsorption density, and volume per site), and the histogram $H(E_i, n_j)$ are calculated by $5 \times 10^6$ MC steps, where the Metropolis algorithm using $P(\bm x)$ is adopted.
Thus, the equilibrium probability distribution reads
\begin{equation}
P_{E,n}^{\beta,\nu}(E_i, n_j) = \frac{H(E_i,n_j) e^{-\beta E_i + \tilde{g}(E_i)}}{\sum_{E_i, n_j}H(E_i,n_j) e^{-\beta E_i + \tilde{g}(E_i)}},
\end{equation}
where the chemical potential term is included in $E_i$, as shown in Eq.~\ref{eq:hamiltonian}. The marginal distributions of the energy and adsorption fraction read
\begin{align}
P_E^{\beta,\nu}(E_i) = \sum_{n_j} P_{E,n}^{\beta,\nu}(E_i, n_j),\\
P_n^{\beta,\nu}(n_j) = \sum_{E_i} P_{E,n}^{\beta,\nu}(E_i, n_j).
\end{align}
The former is presented in Fig.~\ref{fig:physical}{\it C} in the main text. Thermal average of the physical quantities are given by
\begin{equation}
\expval{A}_{\beta,\nu} = \sum_{E_i, n_j} \bar{A}_{E_i, n_j} P_{E,n}^{\beta,\nu}(E_i, n_j).
\end{equation}
They are displayed in Fig.~\ref{fig:physical}{\it A} ($A=N_{\rm ads}/N$), {\it B} ($A=V/N$), and the inset of {\it C} ($A=E/N$) in the main text. The osmotic grand-potential landscape, presented in Fig.~\ref{fig:physical}{\it D} in the main text, can be obtained as follows:
\begin{equation}
\Delta \Omega(n_j) =  -k_{\rm B}T[\log P_n^{\beta,\nu}(n_j)-\min_{n_j}\log P^{\beta,\nu}_n(n_j)].
\end{equation}
We also calculate the energy landscape against the adsorption fraction to decompose the osmotic grand potential into energetic and entropic terms, as shown in Fig.~\ref{fig:supple_landscape}. The conditional probability formula yields
\begin{equation}
E(n_j) = \frac{\sum_{E_i} \bar{E}_{E_i, n_j} P_{E,n}^{\beta,\nu}(E_i,n_j)}{P_{n}^{\beta,\nu} (n_j)}.
\end{equation}
We perform five independent runs and evaluate the statistical errors. However, the standard errors are negligible. Thus, they are omitted from the figure.

\subsection*{Simulations of the conserved adsorption-fraction system}
To examine the role of elastic heterogeneity more deeply, we also perform simulations with a fixed adsorption fraction $N_{\rm ads}/N$ (see \blue{{\it Appendix}, Fig.~S\ref{fig:nads_fix}}). Since the number of adsorbed particles is conserved, we replace the Metropolis sweep for guest adsorptions with the Kawasaki dynamics in the unit MC step~\cite{Landau-Binder}. 
Two nearest neighboring plaquettes $\square$ and $\square'$ are randomly selected, and then a trial exchange of $\sigma_\square$ and $\sigma_{\square'}$ is evaluated under the Metropolis rule.


\subsection*{Acknowledgments}
This work was supported by the JSPS KAKENHI Grant No. JP20H05619.

\bibliography{ref-PCPMOF2}

\clearpage

\renewcommand{\figurename}{{\bf Fig. S}}
\setcounter{figure}{0}

\begin{figure*}[t!]
\centering
\includegraphics[width=14cm]{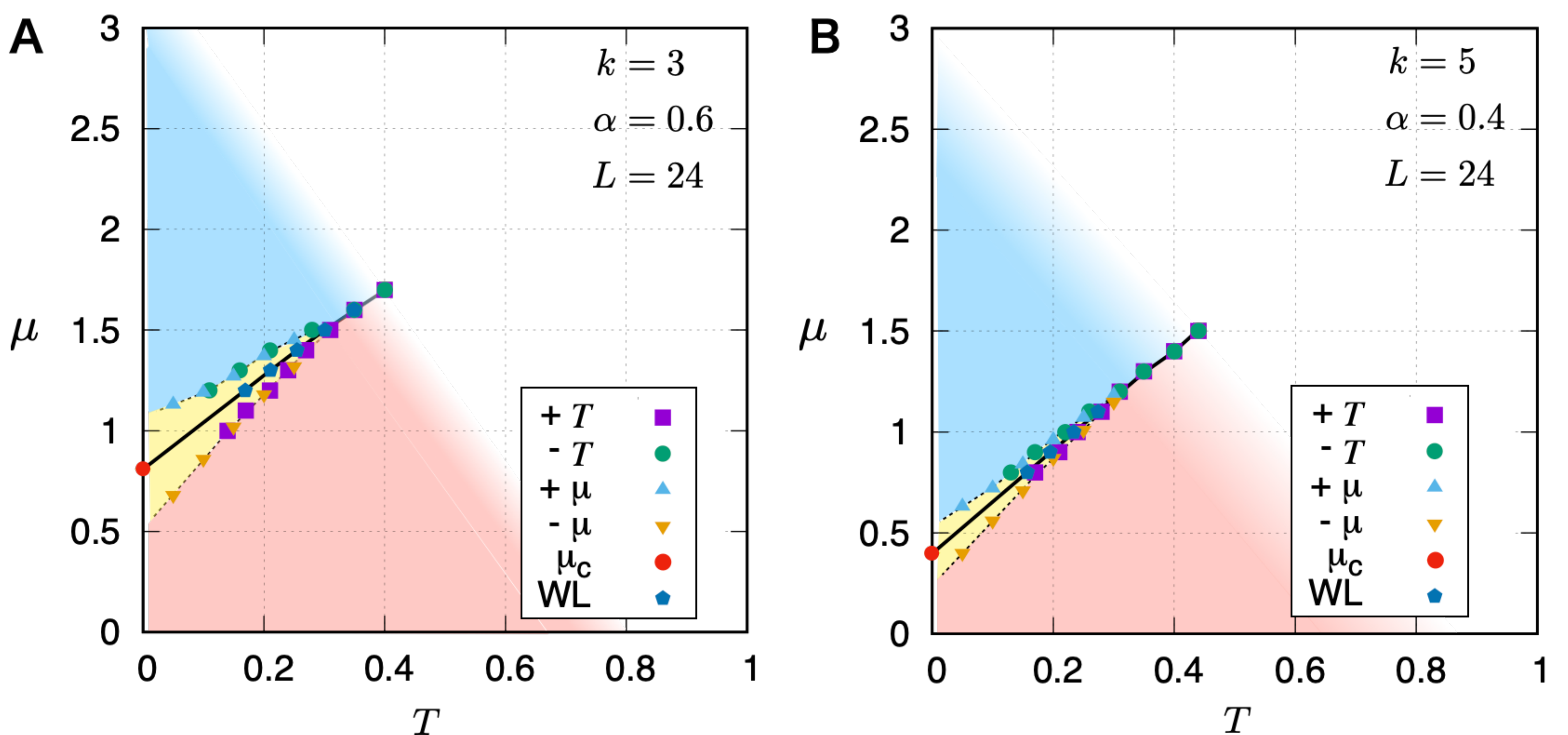}
\caption{{\bf Phase diagram.}
Phase diagram for $k=3$, $\alpha=0.6$ ({\bf A}) and $k=5$, $\alpha=0.4$ ({\bf B}) with $L=24$. The notations are the same as Fig.~\ref{fig:phase} in the main text. The solid curve represents the equilibrium phase boundary between the adsorbed and desorbed phases, determined from the specific-heat peaks obtained by the Wang-Landau (WL) method, whereas the transition point $\mu_{\rm c}=3k\alpha^2/(1+k)$ at $T=0$ is determined analytically by comparing the minimum energy of $V_1$ and $V_1+V_2-\mu$. The boundaries between different colors are determined from the specific-heat peaks obtained by quasi-equilibrium protocols; heating $(+T)$, cooling $(-T)$, increasing chemical potential $(+\mu)$, and decreasing chemical potential $(-\mu)$. The hysteresis region shrinks for weaker elastic heterogeneity (smaller $k$ and $\alpha$).
}
\label{fig:supple_diagram}
\end{figure*}

\begin{figure*}[t!]
\centering
\includegraphics[width=14cm]{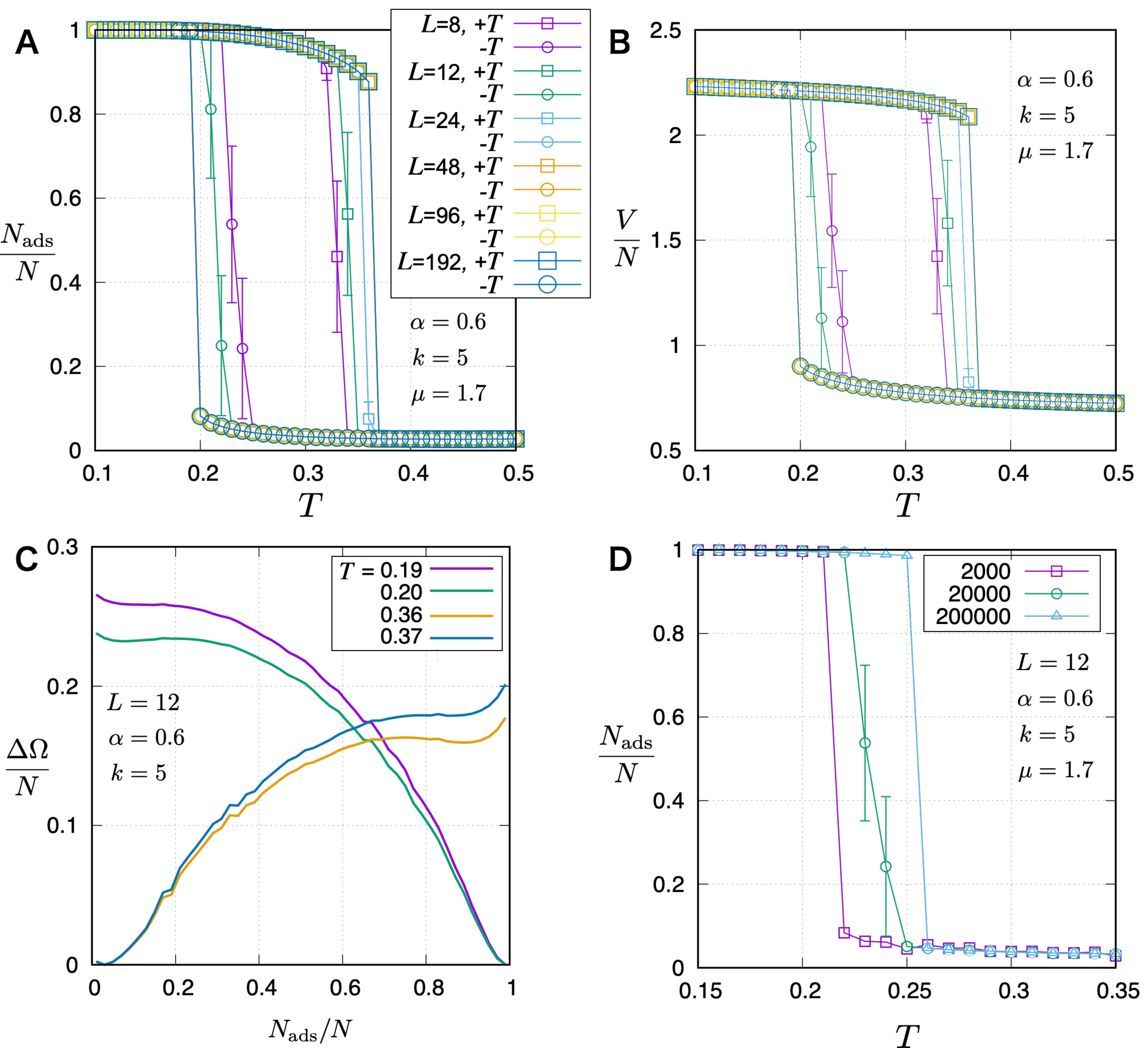}
\caption{{\bf Size dependencies of adsorption hysteresis.}
({\bf A} and {\bf B}) The size dependencies of the adsorption fraction $N_{\rm ads}/N$ (A) and the volume of the system per site $V/N$ (B) obtained by heating $(+T)$ and cooling $(-T)$ protocols at $\mu = 1.7$. Equilibration and sampling Monte Carlo steps $\tau=20000$ at each temperature. ({\bf C}) The free-energy landscape with respect to $N_{\rm ads}/N$ around the spinodal points for $L=12$. Although the transition temperature strongly depends on the system size, the spinodal point does not change. ({\bf D}) The Monte Carlo steps dependency of $N_{\rm ads}/N$ is displayed in the cooling protocol, where the equilibration and sampling Monte Carlo steps $\tau = 2000, 20000, 200000$ at each temperature. The transition from the metastable to the stable state occurs above the spinodal point, exhibiting cooling rate dependency.
The error bar denotes the standard error.
}
\label{fig:supple_size}
\end{figure*}

\begin{figure*}[t!]
\centering
\includegraphics[width=14cm]{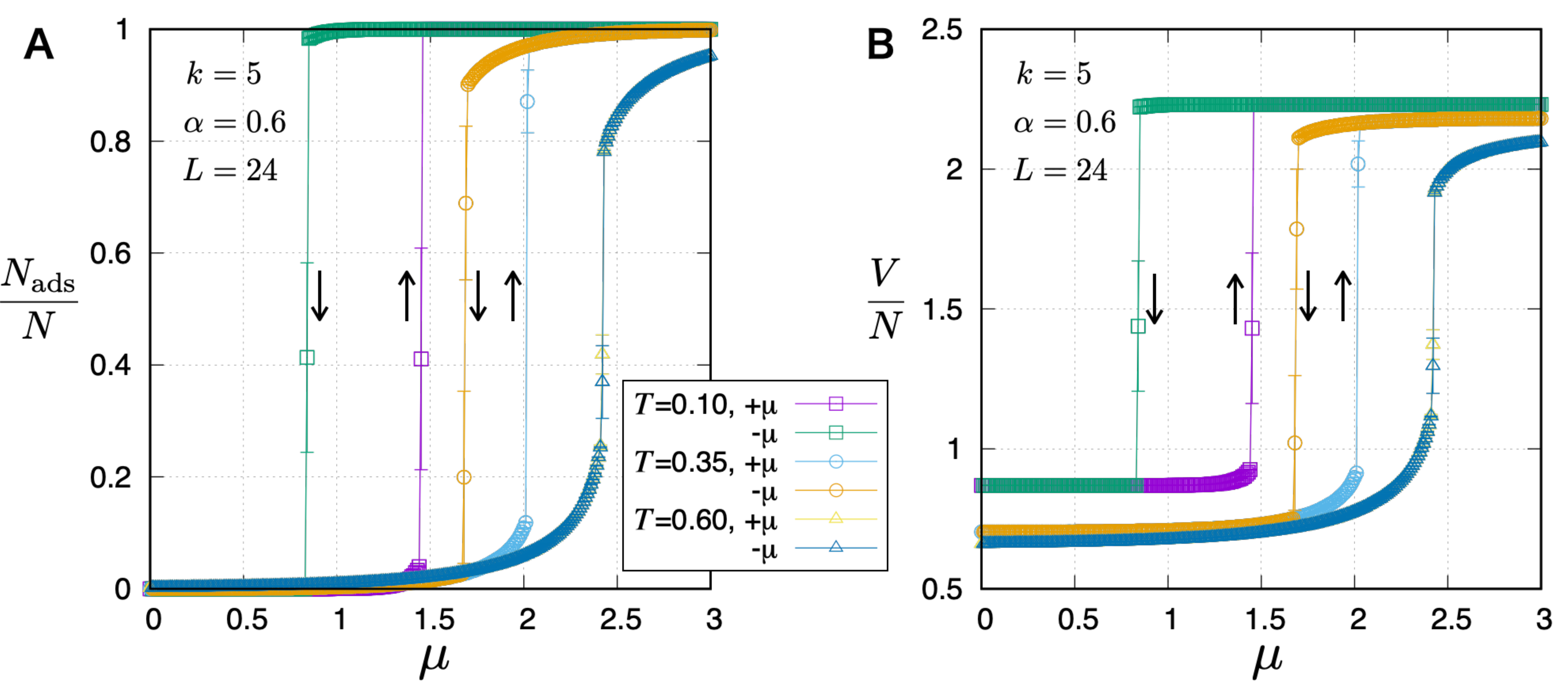}
\caption{{\bf Chemical potential dependencies of physical quantities.}
({\bf A} and {\bf B}) The chemical potential $\mu$ dependencies of the adsorption fraction $N_{\rm ads}/N$ and the volume per site $V/N$ obtained by increasing $\mu$ ($+\mu$) and decreasing $\mu$ ($-\mu$) protocols at $T=0.10, 0.35, 0.60$ for $k=5, \alpha = 0.6$, and $L=24$.
Similar to the temperature dependencies, the jumps in $N_{\rm ads}/N$ and $V/N$ become smaller and hysteretic loops shrink as ($T$, $\mu$) approaches the critical point. The error bars represent the standard error.
}
\label{fig:mu}
\end{figure*}

\begin{figure*}[t!]
\centering
\includegraphics[width=14cm]{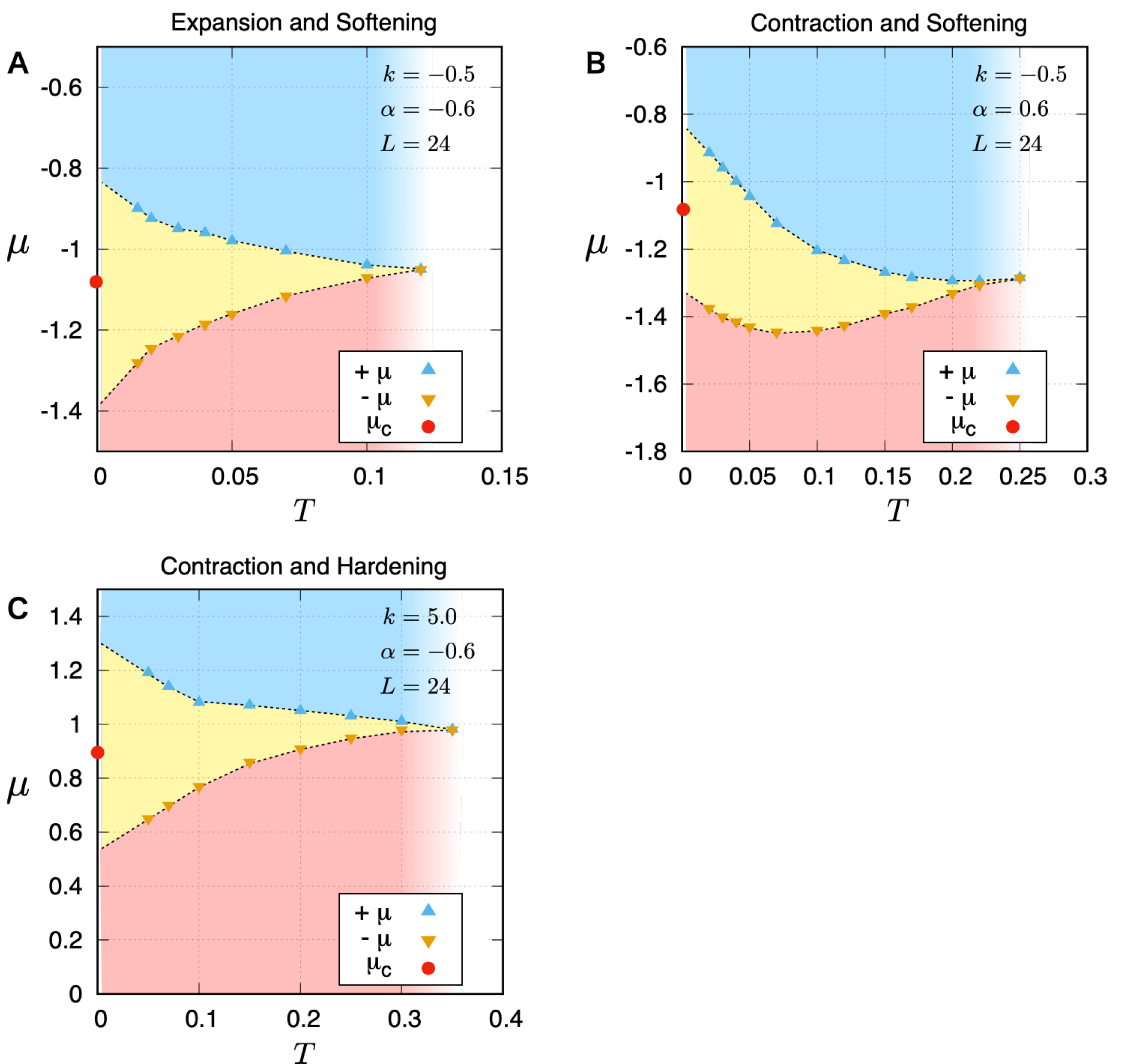}
\caption{{\bf Phase diagram for different signs of $k$ and $\alpha$.}
({\bf A}-{\bf C}) Phase diagram for adsorption expansion and softening condition $k=-0.5$, $\alpha=-0.6$ (A), adsorption contraction and softening condition $k=-0.5$, $\alpha=0.6$ (B) and adsorption contraction and hardening condition $k=5$, $\alpha=-0.6$ (C) with $L=24$. The equilibrium transition point $\mu_{\rm c}=3k\alpha^2/(1+k)$ at $T=0$ is determined analytically by comparing the minimum energy of $V_1$ and $V_1+V_2-\mu$. The shape of the phase boundary changes from the adsorption contraction and hardening condition in Fig.2 in the main text. However, hysteretic behavior due to elastic heterogeneity is conspicuous in all cases. Although we do not obtain equilibrium phase boundaries, it is indicated by the transition points that the slope of the phase boundary is almost zero in (A) and (C) and is negative in (B), implying that the entropy difference is almost zero in (A) and (C), and is negative in (B) according to the Clausius-Clapeyron relation described in the main text.
}
\label{fig:supple_diagram_kalpha}
\end{figure*}

\begin{figure*}[t!]
\centering
\includegraphics[width=14cm]{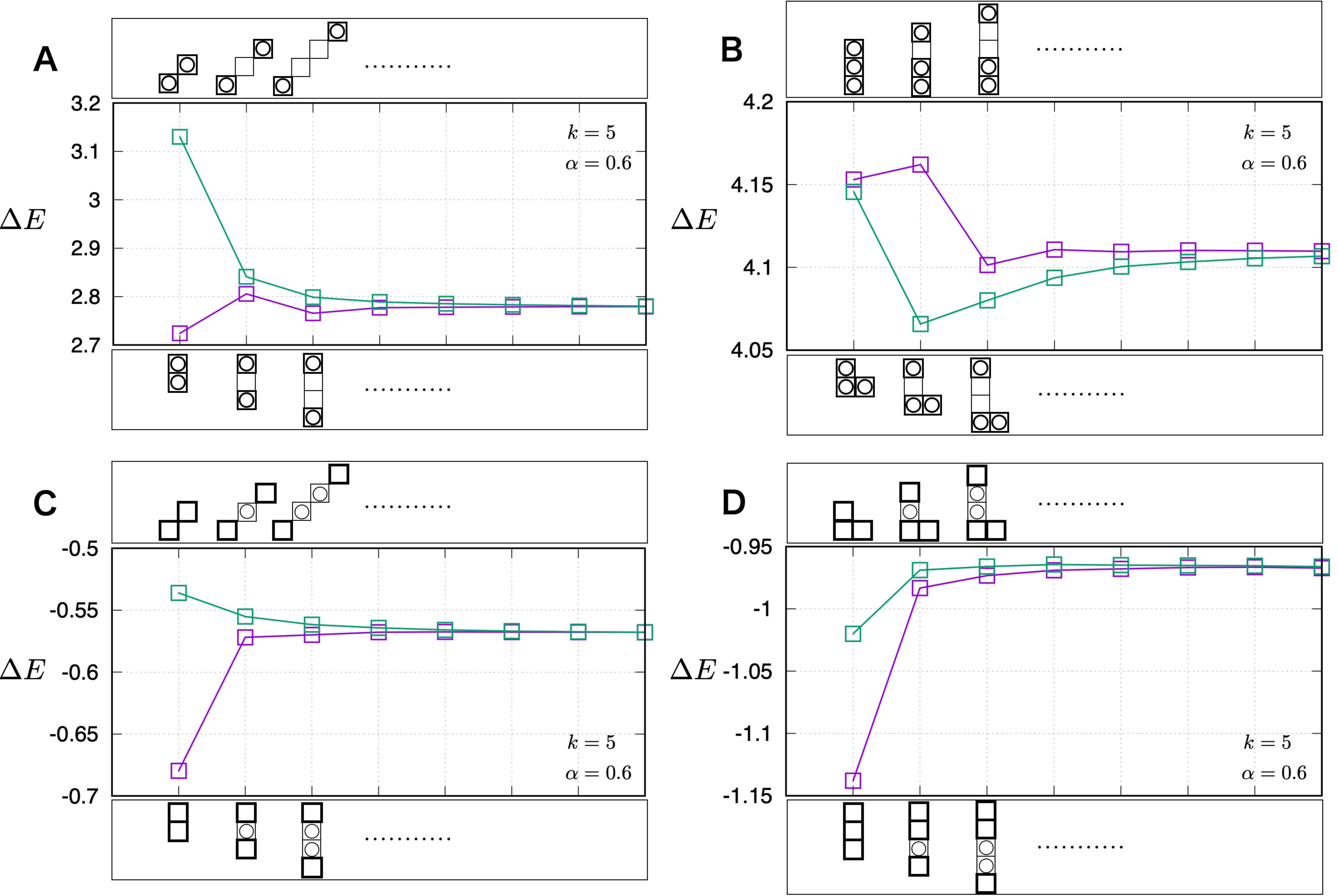}
\caption{{\bf Elasticity-mediated interaction between adsorbed/desorbed sites.}
({\bf A}) The ground state energy of the selected configurations of adsorbed sites in the desorbed matrix. Circles denote the adsorbates, and an open square denotes the desorbed site. $\Delta E$ represents the energy difference between the given configurations and the fully desorbed state. When two adsorbates approach, $\Delta E$ strongly depends on the relative orientation between the adsorbates. In the  horizontally and vertically aligned cases (the lower curve), $\Delta E$ becomes minimum when the dimer is formed. In contrast, two adsorbates repel each other in the diagonally aligned case (the upper curve).
({\bf B}) When a dimer and a monomer approach, $\Delta E$ increase when the trimer is formed. Alternatively, the dimer and the monomer sharing one desorbed site has the minimum energy, indicated by the second configuration in the lower.
({\bf C}) The ground state energy of selected configurations of desorbed sites in the adsorbed matrix. The dimer configuration where two desorbed sites align horizontally and vertically has the lowest energy.
({\bf D}) When a dimer and a monomer of desorbed sites approach, $\Delta E$ becomes lowest when the straight trimer is formed.
$(T,\mu)=(0,0)$, $k=5$, $\alpha=0.6$, and $L=24$ in the figure.
}
\label{fig:supple_effective}
\end{figure*}


\begin{figure*}[t!]
\centering
\includegraphics[width=14cm]{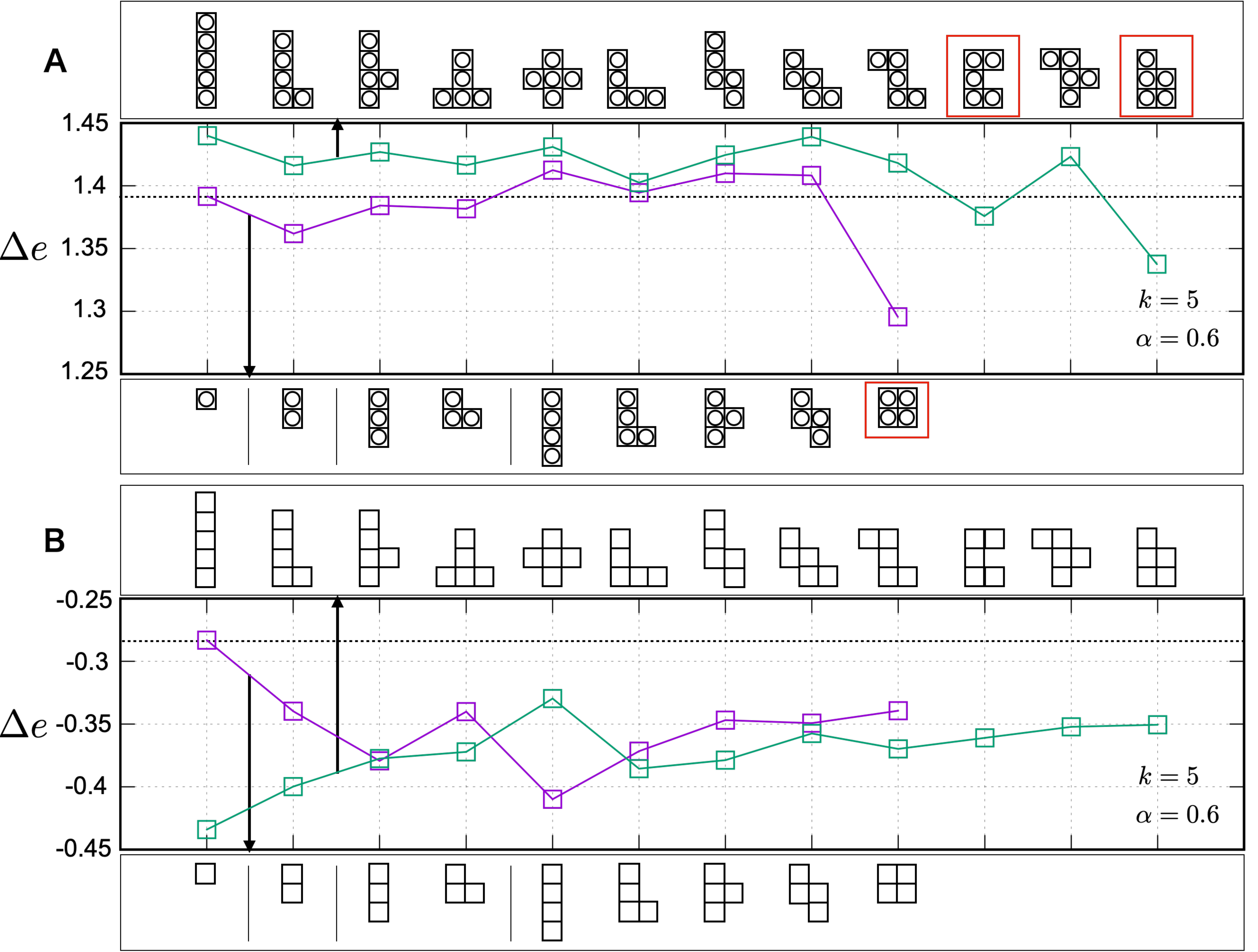}
\caption{{\bf Elastic energy of inclusions.}
We calculate the elastic energy of the adsorbed/desorbed domains with fixed shape in the desorbed/adsorbed matrix at $(T,\mu)=(0,0)$.
({\bf A}) Elastic energy of the adsorbed domains with their size up to five, $\Delta e=\Delta E/N_{\rm ads}$, where $\Delta E$ is the energy difference between the given configuration and the fully desorbed state. Dimer and trimer have lower $\Delta e$ than that of a monomer. However, the tetramer and pentamer have larger elastic energy except for the shapes indicated by the red square, representing the isotropic domain shapes. This result contrast with the liquid-liquid phase separation, wherein $\Delta e$ always becomes smaller in the larger domains than in the monomer by reducing the interfacial area.
({\bf B}) Elastic energy of desorbed domains with their size up to five, $\Delta e = \Delta E/(N-N_{\rm ads})$, where $\Delta E$ is the energy difference between the given configuration and the fully adsorbed state. In contrast to (A), $\Delta e$ always becomes lower in the larger domains than in the monomer. However, the straight-shape domain, not the isotropic domains, has the lowest $\Delta e$ among the same cluster size and becomes lower as the domain grows.
$k=5$, $\alpha=0.6$, and $L=24$.
}
\label{fig:domain_size}
\end{figure*}

\begin{figure*}[t!]
\centering
\includegraphics[width=12cm]{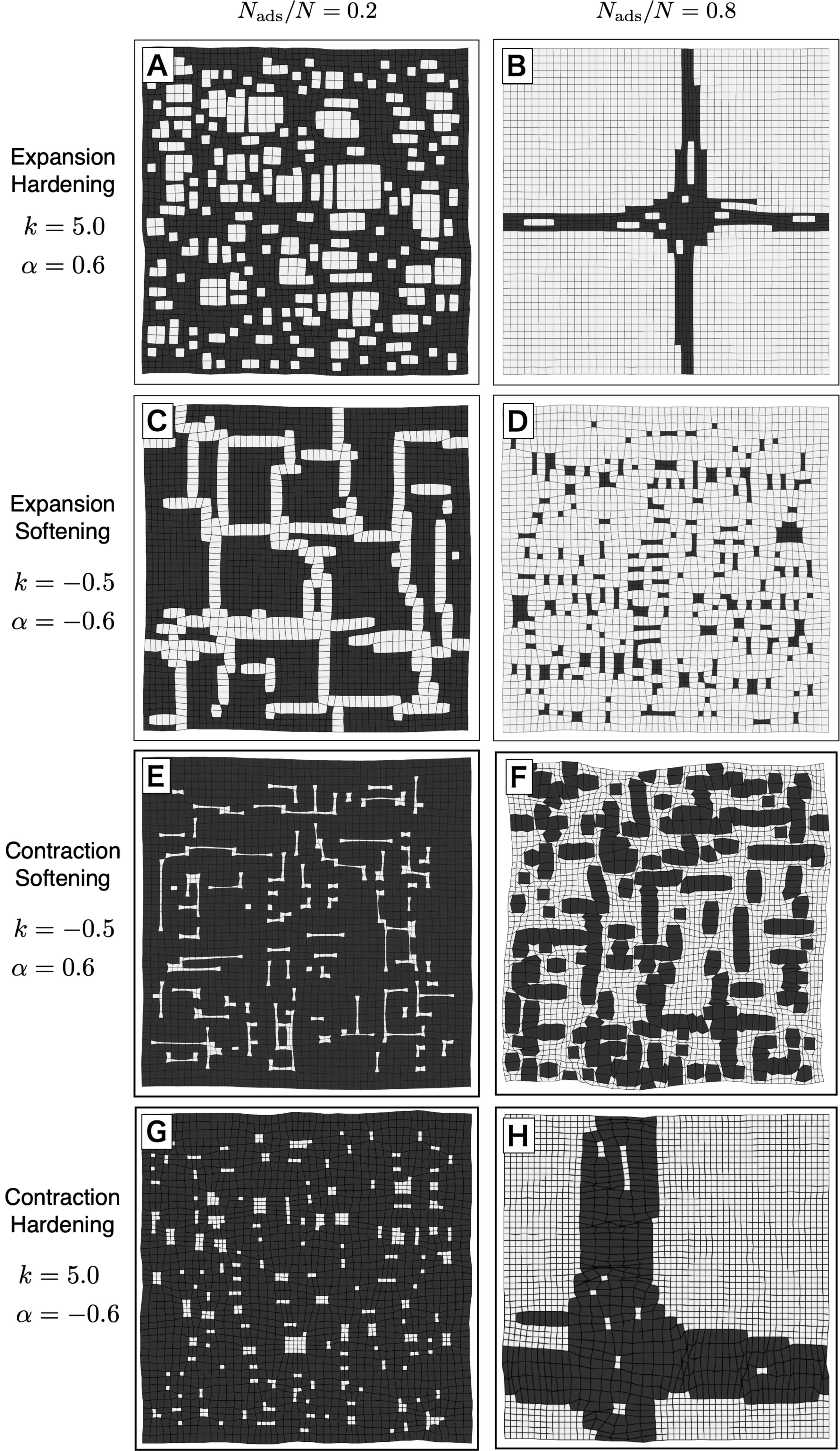}
\caption{{\bf Spatial heterogeneity.}
Distribution of the adsorbed (white) and desorbed (black) sites under fixed $N_{\rm ads}/N$ simulations. The systems are annealed to $T=0.01$ from high temperature.
({\bf A} and {\bf B}) Snapshots under adsorption expansion and hardening condition ($k=5$, $\alpha=0.6$). The harder adsorbed domains have isotropic shapes and are dispersed in the softer desorbed matrix when $N_{\rm ads}/N=0.2$ (A). In contrast, a single softer desorbed domain is percolated in the harder adsorbed matrix when $N_{\rm ads}/N=0.8$ (B).
({\bf C} and {\bf D}) Snapshots under adsorption expansion and softening condition. In contrast to the adsorption hardening condition ($k=-0.5$ and $\alpha=-0.6$), the softer adsorbed domain percolates in the desorbed phase (C), whereas the harder desorbed domains disperse in the softer adsorbed matrix in the adsorbed phase (D).
({\bf E} and {\bf F}) Snapshots under adsorption contraction and softening condition ($k=-0.5$ and $\alpha=0.6$). The adsorbed domains in (E) form more anisotropic shapes than the desorbed domains in (F).
({\bf G} and {\bf H}) Snapshots under adsorption contraction and hardening condition ($k=5$ and $\alpha=-0.6$).
The behavior is similar to one of the adsorption expansion and hardening condition (A and B).
}
\label{fig:nads_fix}
\end{figure*}

\begin{figure*}[t!]
\centering
\includegraphics[width=14cm]{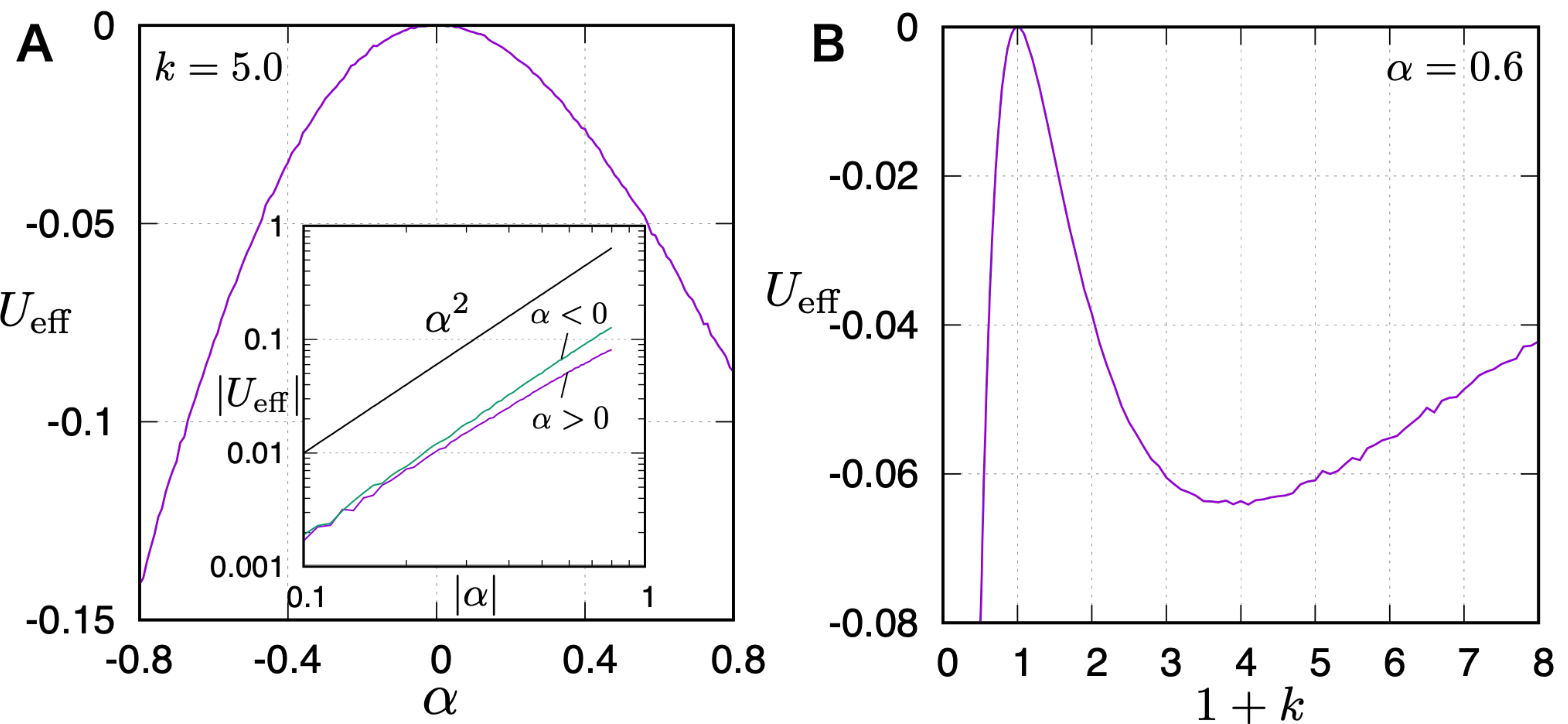}
\caption{{\bf $k$ and $\alpha$ dependency of the effective interaction between guest particles.}
We calculate the elastic energy difference between an adsorption dimer and two completely isolated adsorption at $(T,\mu) = (0, 0)$.
$\alpha$ dependency with $k=5.0$ ({\bf A}) and $k$ dependency with $\alpha = 0.6$ ({\bf B}).
The interaction energy $U_{\rm eff}$ is proportional to $\alpha^2$ (A) and behaves non-monotonically with respect to $k$ (B).
}
\label{fig:supple_Ueff}
\end{figure*}

\begin{figure*}[t!]
\centering
\includegraphics[width=14cm]{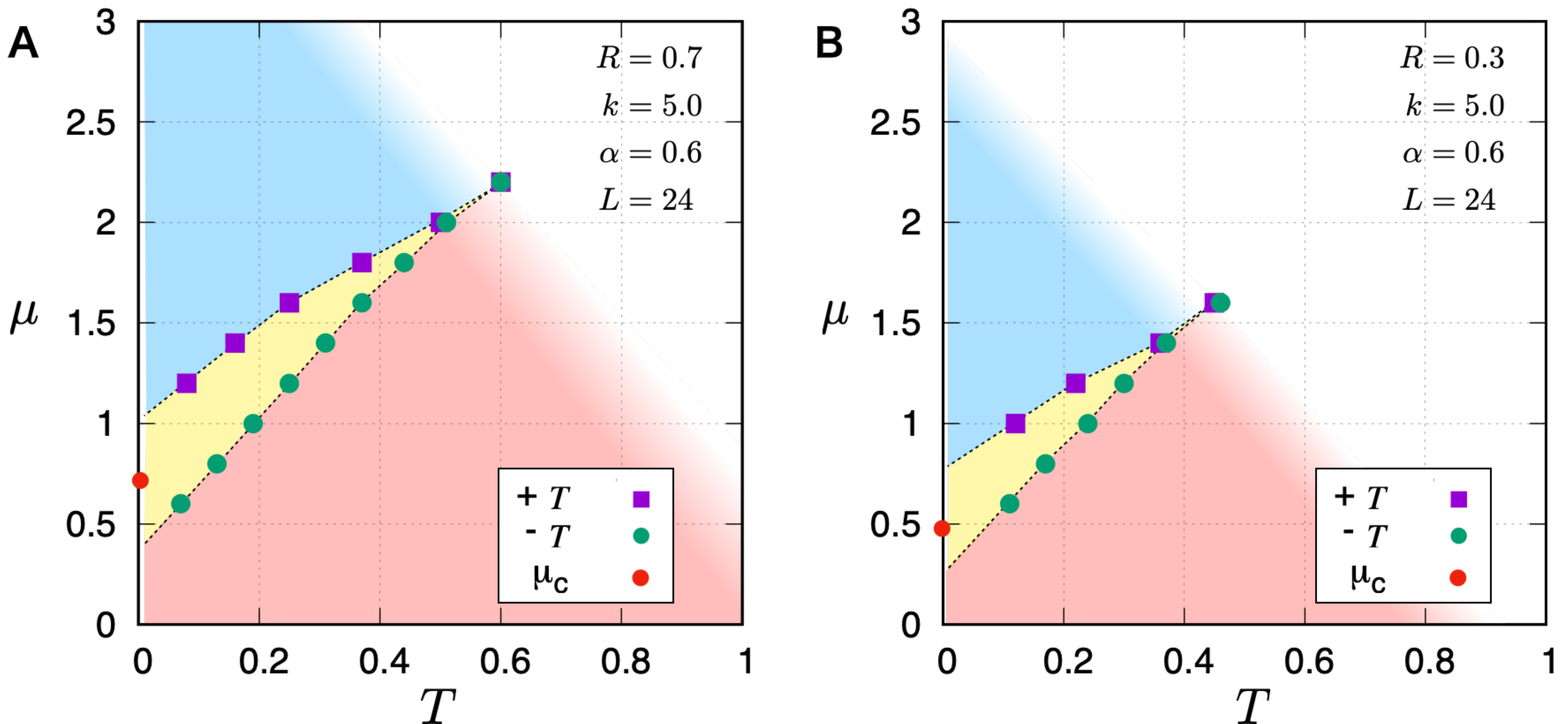}
\caption{{\bf Phase diagram for weaker next nearest interactions.}
({\bf A} and {\bf B}) Phase diagram for $k=5$, $\alpha=0.6$, $L=24$ with $R=0.7$ ({\bf A}) and $R=0.3$ ({\bf B}), where $R$ is the ratio of the next-nearest-neighbor interaction to the nearest-neighbor interaction. The equilibrium transition point $\mu_{\rm c}=(1+2R)k\alpha^2/(1+k)$ at $T=0$ is determined analytically by comparing the minimum energy of $V_1$ and $V_1+V_2-\mu$. The qualitative feature remains unchanged by varying $R$, though the transition temperatures and chemical potentials decrease, and hysteresis shrinks as $R$ becomes small.
}
\label{fig:supple_diagram_NNN}
\end{figure*}

\begin{figure*}[t!]
\centering
\includegraphics[width=14cm]{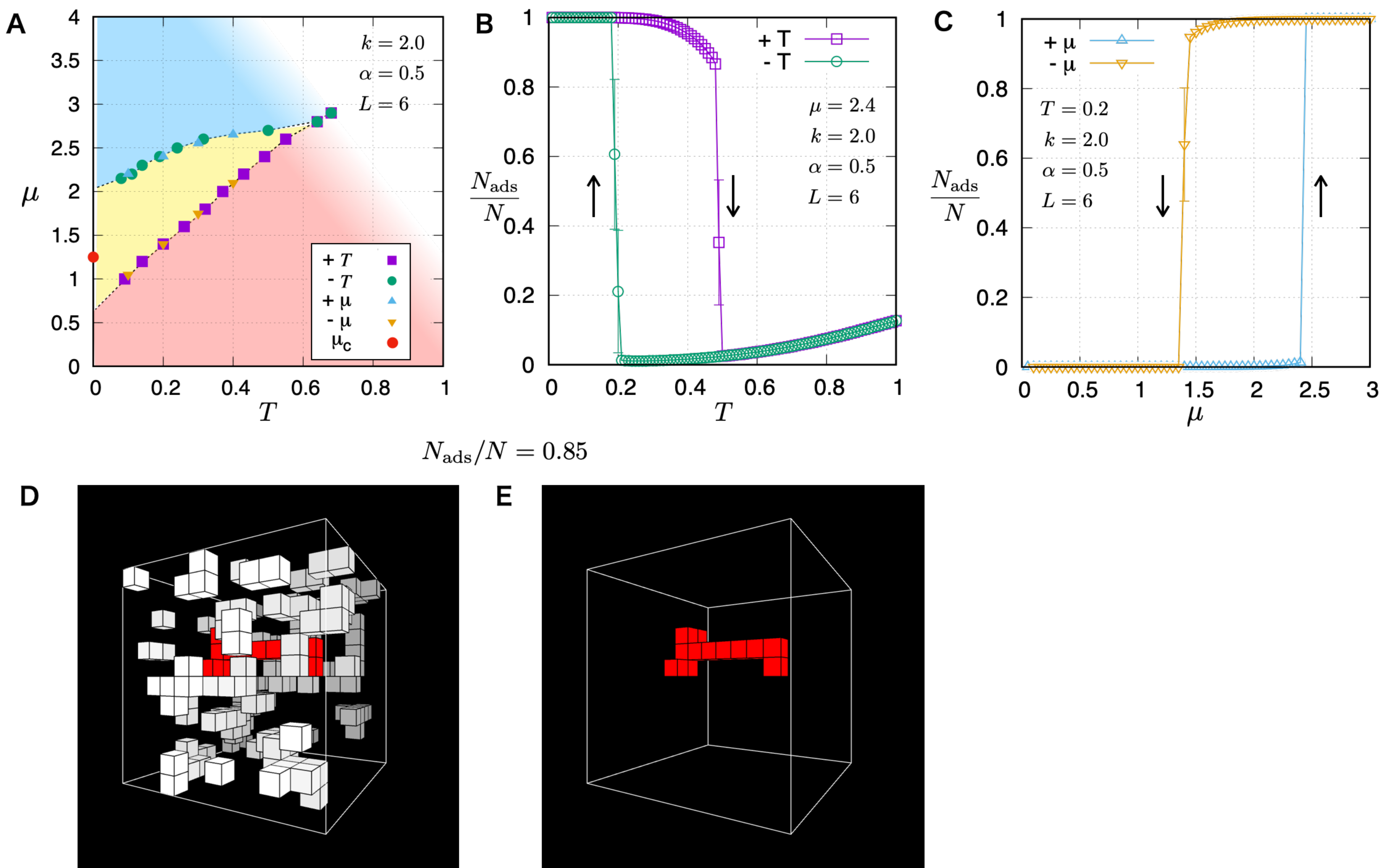}
\caption{{\bf Results for a three-dimensional model on a simple cubic lattice.}
({\bf A}) Phase diagram for $k=2$, $\alpha=0.5$ with $L=6$. The adsorption-desorption hysteresis is qualitatively the same as the two-dimensional result. ({\bf B} and {\bf C}) The adsorption fraction at $\mu = 2.4$ by increasing and decreasing the temperature (B) and $T = 0.2$ by increasing and decreasing the chemical potential (C). ({\bf D} and {\bf E}) Distribution of the desorbed sites under fixed $N_{\rm ads}/N$ simulations. The systems are annealed to $T=0.01$ from high temperature. The maximum size domain is displayed in red, as depicted in (E). The domains of the desorbed sites forms anisotropic shapes: the boundary surface area per site $\gamma=4.17$ for the maximum size (red) domain. In the three-dimensional model on a simple cubic lattice, the potential energy reads $V_1(\{\bm{r}_{i \in \square} \})=\frac{1}{8}\sum_{\rm NN}(1-r_{ij})^2 + \frac{1}{4}\sum_{\rm NNN}(\sqrt{2}-r_{ij})^2$ and $V_2(\{\bm{r}_{i \in \square} \})=k[\frac{1}{8}\sum_{\rm NN}(1+\alpha-r_{ij})^2 + \frac{1}{4}\sum_{\rm NNN}(\sqrt{2}(1+\alpha)-r_{ij})^2]$.
}
\label{fig:supple_3d}
\end{figure*}

\end{document}